\documentclass[12pt]{article}
\usepackage{a4wide,epsfig,amsmath,amssymb,cite,scalefnt,wasysym}
\usepackage{color}
\usepackage{makeidx}
\usepackage{array}
\usepackage{longtable}
\usepackage{textcomp}
\usepackage{slashed}
\usepackage{listings}

\definecolor{Black}{RGB}{0,0,0}
\definecolor{SkyBlue}{RGB}{200,225,255}

\newcommand{\gsim}{\;\rlap{\lower 3.5 pt \hbox{$\mathchar \sim$}} \raise 1pt
 \hbox {$>$}\;}
\newcommand{\lsim}{\;\rlap{\lower 3.5 pt \hbox{$\mathchar \sim$}} \raise 1pt
 \hbox {$<$}\;}

\newcommand*{\vcenteredhbox}[1]{\begingroup
\setbox0=\hbox{#1}\parbox{\wd0}{\box0}\endgroup}

\allowdisplaybreaks

\makeindex

\begin{document}

\title{\vskip-3cm{\baselineskip14pt
    \begin{flushleft}
      \normalsize ALBERTA-THY-10-16
  \end{flushleft}}
  \vskip1.5cm
  Irreducible tensor basis and general Fierz relations for\\
  Bhabha scattering like amplitudes 
}

\author{
  Tao Liu$^a$ and Nikolai Zerf$^b$
  \\[1em]
 $~^a${\small\it Department of Physics}\\
 {\small\it University of Alberta}\\
 {\small\it Edmonton AB T6G 2J1, Canada}\\[0.3cm]
 $~^b${\small\it Institut f{\"u}r Theoretische Physik}\\
 {\small\it Universit\"at Heidelberg}\\
 {\small\it  D-69120 Heidelberg, Germany}
}

\date{}

\maketitle

\thispagestyle{empty}

\begin{abstract}
 \medskip
  \noindent
We construct an irreducible s- and t-channel tensor basis for Bhabha scattering like amplitudes based on the properties of the underlying Lorentz symmetry in four space-time dimensions.
In the given basis the calculation of amplitude contractions like the amplitude square reduces to the contraction of their corresponding coefficient tensors.
Further the basis retains the full amplitude information and thus can be applied in off-shell cases.
The general Fierz transformations which relate the s- and t-channel basis with each other are obtained.
As an example for application we use the basis to calculate the tree-level Bhabha scattering amplitude.
\end{abstract}

\thispagestyle{empty}


\newpage

\section*{Introduction}
In high energy physics the calculation of S-matrix elements or scattering amplitudes within Quantum Field Theory are required to obtain predictions for physical observables like cross section and decay rates.
Although the amplitude corresponding to a certain scattering process contains the full physical information, 
one has to calculate its square or absolute value in order to determine the transition probability.
Depending on the order of perturbation theory within which the amplitude is to be calculated the evaluation of the amplitude 
square can be computationally expensive.
One reason for this is that the number of terms in the amplitude square grows quadratically with the number of terms in the amplitude using a naive squaring algorithm.
However, because amplitudes in non-Lorentz violating theories have to transform in a Lorentz covariant way
one can use the underlying Lorentz symmetry to decompose any truncated amplitude into irreducible orthogonal basis tensors.
This immediately restricts the squaring of an amplitude to each irreducible sub-block.
In case where one is only interested in the interference of, for example, tree-level amplitudes with amplitudes of higher order, 
this can thus greatly speed up the calculation of relevant terms, because only the sub-blocks appearing in the decomposition of the tree-level amplitude can contribute.
One of course does not need any irreducible Lorentz tensor basis to be able to calculate radiative corrections to cross section,
because one can just calculate the unpolarized forward scattering amplitude allowing to get rid of all tensor structures including fermion lines.
However, in this case the full amplitude information is lost and one would need to recalculate the amplitude again if any other type of interference is required,
although the full amplitude calculation would not pose any further technical difficulties.
Further the number of Feynman diagrams corresponding to a forward scattering amplitude indeed grows with the number of diagrams $n$ of the process amplitude like $n^2$.
On top of that the calculation of each forward scattering diagram is clearly more involved than the calculation of the corresponding process diagram.

It is thus desirable to construct an irreducible Lorentz tensor basis, which allows to decompose any truncated amplitude in this basis.
This reduces the problem of the amplitude calculation to the determination of the coefficients of the tensor basis involving only Lorentz scalar quantities during the evaluation.
In this article we follow this idea and construct such a basis for Bhabha scattering like processes in four space-time dimensions.
We construct an s-channel and t-channel basis which is independent of the specific kinematics and thus can be used in the off-shell case, too.
Further we show how amplitudes in the s-channel or t-channel basis can be related to each other via general Fierz transformations.
As an example we apply the basis to the calculation of the tree-level Bhabha scattering amplitude.

This article is organized as follows.
In the first section we introduce our notation for the basic invariant tensors of the Lorentz group $SO(3,1)$ and useful relations amongst them.
Next we generically construct the tensor basis for Bhabha scattering like amplitudes in the s-channel setup.
In the third we provide all explicit expressions which are required for a specific realization of our generic tensor bases in $d=4$ space-time dimensions. 
In Sect.~4 we discuss how to relate amplitudes given in the $s$-channel basis to the amplitude in the $t$-channel basis via general Fierz transformation.
In the fifth section we use the tensor basis to evaluate generically squares of amplitudes leading to a simple contraction prescription of the relevant coefficient tensors.
In the last section we apply our method to the tree-level Bhabha scattering amplitude.

In the Appendices we provide a discussion about the application of Schouten identities 
and information which enables the reader to apply our methods using the electronic files (including examples) provided with this publication without any further work.   

\section{Basic tensors}
Throughout this article we are working with Lorentz tensors.
The components of the four-dimensional vector/spinor representation are labeled by indices $\mu,\nu,\sigma,\dots$/$\alpha,\beta,\dots$.
The dimension of the representation is given by $d=4$.
Because the symmetry of the Lorentz group is of type $SO(3,1)$ we have a symmetric invariant tensor of rank two, the metric $g^{\mu \nu}$,
and a fully antisymmetric Levi-Civita tensor $\epsilon^{\mu \nu \rho \sigma}$ of rank four.
In the spinor space we have an invariant rank two tensor, the Kronecker delta $\delta_{\alpha}^{\beta}$ which enables to write down invariant complex scalar products.
The dimension of the spinor representation is given by $d_s=4$.
Further we have four four-dimensional Dirac matrices $(\gamma^{\mu})_{\alpha}^{\beta}$ which couple the fundamental and the anti-fundamental spinor representation to the vector representation.
The $\gamma$ obey the Clifford algebra:
\begin{align}
\{\gamma^{\mu},\gamma^{\nu}\}=2 g^{\mu \nu} \mathbb{I}_{d_s\times d_s}\,.
\end{align}
In explicit spinor components we have $(\mathbb{I}_{d_s\times d_s})_{\alpha}^{\beta}=\delta_{\alpha}^{\beta}$.

\begin{figure}
\centering
\begin{tabular}{llll}
 \vcenteredhbox{$\delta\,=$} & \vcenteredhbox{\includegraphics[scale=0.6]{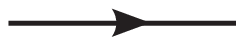}} \vcenteredhbox{$\,,$} & \vcenteredhbox{$\tilde{\varepsilon}\,=$} & \vcenteredhbox{\includegraphics[scale=0.6]{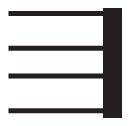}}\vcenteredhbox{$\,,$} \\
 \vcenteredhbox{$g\,=$ } &  \vcenteredhbox{\includegraphics[scale=0.6]{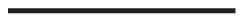}} \vcenteredhbox{$\,,$} & \vcenteredhbox{$\Gamma_{i}\,=$} & \vcenteredhbox{\includegraphics[scale=0.6]{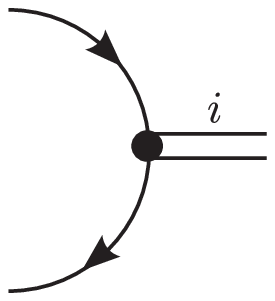}}\vcenteredhbox{$\,.$} \\
\end{tabular}
 \caption{Bird track notation for the basic invariant tensors of ${\rm SO}(3,1)$\label{FIG:BasicInvariants}.}
\end{figure}
All operators acting on the spinor space can be written as linear combinations of 16 independent bilinears $\Gamma$.
We use the following shorthand for them:
\begin{align}
 \Gamma_{\underline{0}} & = \mathbb{I}_{4\times 4}\,,& \Gamma_{\underline{1}} & = \gamma^{\mu}\,, & \Gamma_{\underline{2}} & = \frac{1}{2\sqrt{2}} [\gamma^{\mu},\gamma^{\nu}]\,, \nonumber\\
 \Gamma_{\underline{3}} & =\gamma^{\mu} \gamma^{5}\,,& \Gamma_{\underline{4}} & = \gamma^{5}\,,
\end{align}
where we use 
\begin{align}
 \gamma^{5}=\frac{i}{4!} \varepsilon^{\mu \nu \rho \sigma}\gamma_{\mu}\gamma_{\nu}\gamma_{\rho}\gamma_{\sigma}\,.
\end{align}
In Minkowski space $\varepsilon^{0123}=1=-\varepsilon_{0123}$.
It is convenient to absorb one imaginary unit and a multiplicity factor into the definition of the symbol 
\begin{align}
 \tilde{\varepsilon}^{\mu \nu \rho \sigma} &= \frac{i}{\sqrt{4!}}\varepsilon^{\mu \nu \rho \sigma} \,.
\end{align}
This leads to the contraction identity:
\begin{align}
 \tilde{\varepsilon}^{\mu \nu \rho \sigma}\tilde{\varepsilon}_{\mu \nu \rho \sigma} &= 1\,.
\end{align}
The Dirac bilinear index $i$ ($j,k,l$) labeling the $\Gamma_{i}$ counts the number of anti-symmetric vector representation indices and thus labels the irreducible representation (irrep)
with dimension:
\begin{align}
 d_i&={{d_{\underline{1}}}\choose{i}}\,.
\end{align}
With the special case $d_{\underline{1}}=d$.
We underline explicit values for Dirac bilinear indices,
in order to not confuse them with explicit representation labels $\mathcal{R}_i$ which will be introduced later.

To be able to give compact expression for tensor contractions in form of diagrams, 
we introduce a bird track notation for each invariant tensor in Fig.~\ref{FIG:BasicInvariants}.
A bird track graph can be understood as group selected component of a Feynman diagram.
Throughout this paper we select the Lorentz part of the Feynman diagram.
A complete introduction to the bird track notation can be found for example in Ref.~\cite{Cvitanovic:2008zz}.

\begin{figure}
\centering
\begin{tabular}{llllll}
 \vcenteredhbox{$d_s\,=$} & \vcenteredhbox{\includegraphics[scale=0.4]{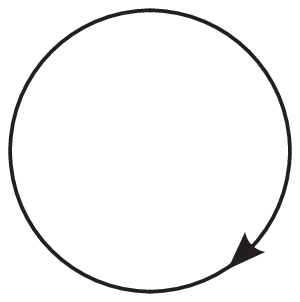}} \vcenteredhbox{$\,,$} & 
 \vcenteredhbox{$d_{\underline{1}}\,=$} & \vcenteredhbox{\includegraphics[scale=0.4]{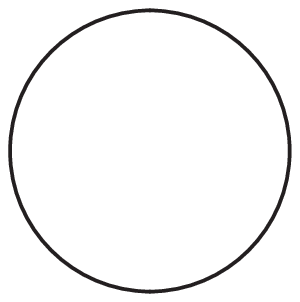}}\vcenteredhbox{$\,,$} & 
 \vcenteredhbox{$d_i\,=$} & \vcenteredhbox{\includegraphics[scale=0.4]{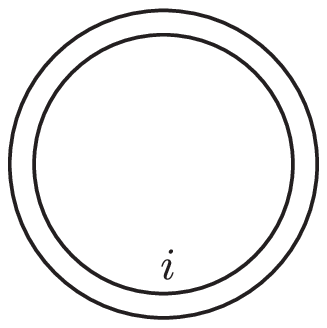}}\vcenteredhbox{$\,.$}\\
\end{tabular}
 \caption{Dimensions of irreps in bird track notation. \label{FIG:IrrepDimensions}}
\end{figure}

As a simple example of the bird track notation we show the dimension of selected representations in Fig.~\ref{FIG:IrrepDimensions}.

\begin{figure}
\centering
\begin{tabular}{cc}
\vcenteredhbox{\includegraphics[scale=0.7]{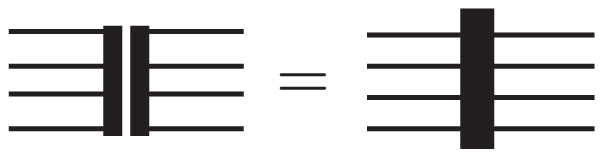}}$\qquad$ & \vcenteredhbox{\includegraphics[scale=0.7]{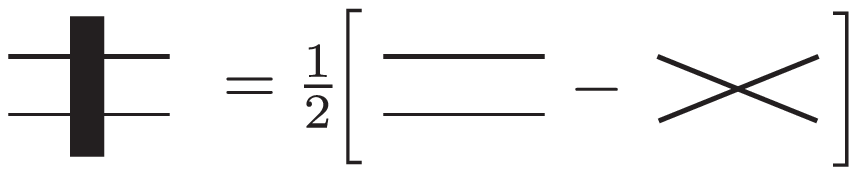}}\\
 (a) &  (b) \\
  \vcenteredhbox{\includegraphics[scale=0.7]{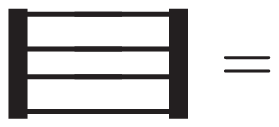}}$\qquad$   &  \vcenteredhbox{\includegraphics[scale=0.5]{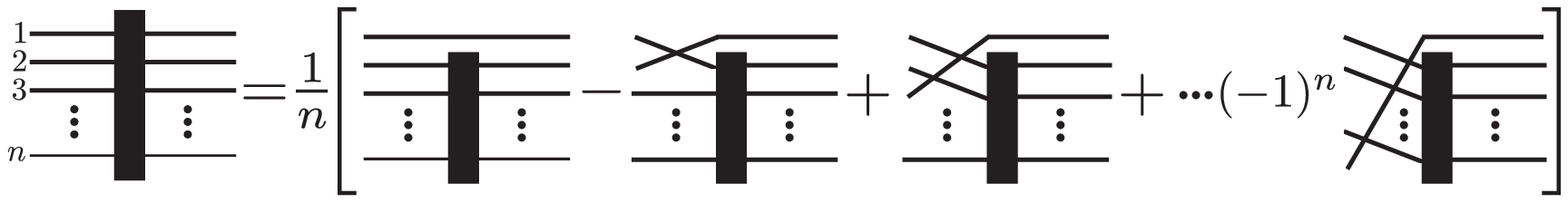}}\\
 (c) &  (d)
\end{tabular}
 \caption{(a) Relation between two Levi-Civita and the symmetrizer for four indices for ${\rm SO}(3,1)$.
          (b) Basic definition of the antisymmetrizer for two indices.        
          (c) Normalization condition of the Levi-Civita tensor\label{FIG:UseFulRelations}. Here the right hand side of the equation is one.
          (d) Recursive definition for antisymmetrizer with $n$ indices. The sign in front of each term is given by the total number of line crossings $n$ via $(-1)^n$.
 }
\end{figure}

\begin{figure}
\centering
\includegraphics[scale=0.7]{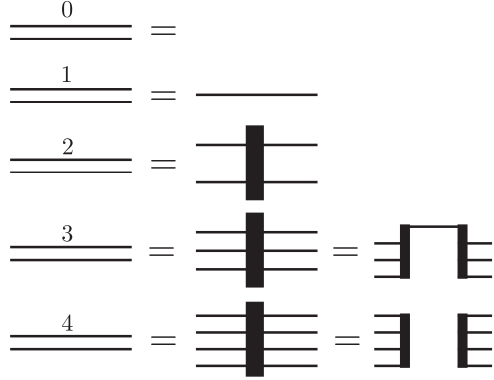}
 \caption{Bird track notation of. Eq.~(\ref{EQ:HigherRepMetricDef}\label{FIG:HigherRepMetric})}
\end{figure}

With the definition of the anti-symmetrizer in bird track notation shown in Fig.~\ref{FIG:UseFulRelations} (b,d) and the relation for a product of two Levi-Civita tensors (a),
we show in Fig.~\ref{FIG:HigherRepMetric} how to construct higher dimensional irreps $i$ using multiple anti-symmetrized vector representations.
Here the application of two Levi-Civita tensors allows to diagonalize the antisymmetric $i={\underline{4}}$ and $i={\underline{3}}$ representation making the dimension of each irrep explicit $1$- and $4$-dimensional.
In the following we will call the $i=\underline{4}$ representation $1^-$ in order to distinguish it from the $i=\underline{0}$ alias $1^+$ representation.
In the same fashion one should not confuse the $i=\underline{3}$ alias $4^-$ with the $i=\underline{1}$ alias $4^+$ irrep.
Further we have the $i=\underline{2}$ alias $6$ which corresponds to the adjoint representation of the Lorentz group.

Because one can always express an even number of $\varepsilon$-tensors via metric tensors using the relation given in Fig.~\ref{FIG:UseFulRelations} (a).
Any tensor either contains zero or one $\varepsilon$-tensor. 
Thus one can attribute each tensor a so called parity $P_{\varepsilon}(T)=(-1)^{n_{\varepsilon}}$, 
where $n_{\varepsilon}$ is the number of $\varepsilon$-tensor hidden inside the respective tensor $T$.
As special cases we have for example $P_{\varepsilon}(\varepsilon) =-1$ and $P_{\varepsilon}(1) =+1$.

The introduced shorthand for the higher dimensional symmetric tensor of rank two within the given irreps can be written in terms of products of the metric tensor in the vector representation like follows:
\begin{align}
 g^{(\underline{0})}_{\{\mu\}\{\nu\}} &=1\,,&  g^{(\underline{1})}_{\{\mu\},\{\nu\}} &=g_{\mu \nu}\,, \nonumber\\
 g^{(\underline{2})}_{\{\mu\}\{\nu\}} &=\frac{1}{2}(g_{\mu_1 \nu_1}g_{\mu_2 \nu_2}-g_{\mu_1 \nu_2}g_{\mu_2 \nu_1})\,, \nonumber\\
 g^{(\underline{3})}_{\{\mu\}\{\nu\}} &=g_{\mu \nu} \,,&  g^{(\underline{4})}_{\{\mu\}\{\nu\}} &=1\,.\label{EQ:HigherRepMetricDef}
\end{align}
Here we choose the convention to absorb all Levi-Civita $\varepsilon$-tensors into the connected Dirac bilinears.
The parity of the corresponding Dirac bilinear allows to distinguish the $1^-$ from the $1^+$ ($4^+,4^-$) representation. 
Here and in the following an index in wavy brackets is understood as collective index.

\begin{figure}
\centering
   \includegraphics[scale=0.6]{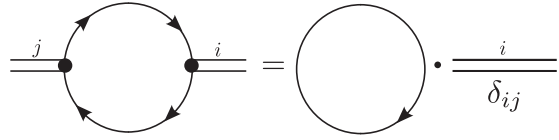}  \\
 \caption{Bird track notation of Eq.~(\ref{EQ:BilinearOrthogonality})\label{FIG:DiracBilinear}.}
\end{figure}

We further introduce the barred version of the 16 bilinears:
\begin{align}
 \overline{\Gamma}_{i} & = \gamma_0\Gamma_i^{\dagger}\gamma_0\,.
\end{align}
with $\gamma_0\gamma^{\dagger}_{\mu}\gamma_0=\gamma_{\mu}$  and $\gamma_0\gamma_5^{\dagger}\gamma_0=-\gamma_{5}$ we find:
\begin{align}
 \overline{\Gamma}_{\underline{0}} & = \Gamma_{\underline{0}} \,,& \overline{\Gamma}_{\underline{1}} & =\Gamma_{\underline{1}} \,, & \overline{\Gamma}_{\underline{2}} & = -\Gamma_{\underline{2}}\,, \nonumber\\
 \overline{\Gamma}_{\underline{3}} & = \Gamma_{\underline{3}} \,,&  \overline{\Gamma}_{\underline{4}}&= -\Gamma_{\underline{4}}\,. & &
\end{align}
It is straight forward to check that the following orthogonally relation is fulfilled: 
\begin{align}
 \sigma_i {\rm tr}\{ \overline{\Gamma}_{i}\Gamma_{j}\} =  \delta_{i j} {\rm tr}(\mathbb{I}_{d_s\times d_s}) g^{(i)}_{\{\mu\}\{\nu\}}\,.\label{EQ:BilinearOrthogonality}
\end{align}
Here $\sigma_i$ is a sign which is plus for $0\leq 2$ and minus for  $i>2$.
The above equation can be cast into bird track notation like shown in Fig.~\ref{FIG:DiracBilinear}.

\begin{figure}
\centering
\begin{tabular}{cc}
\vcenteredhbox{\includegraphics[scale=0.6]{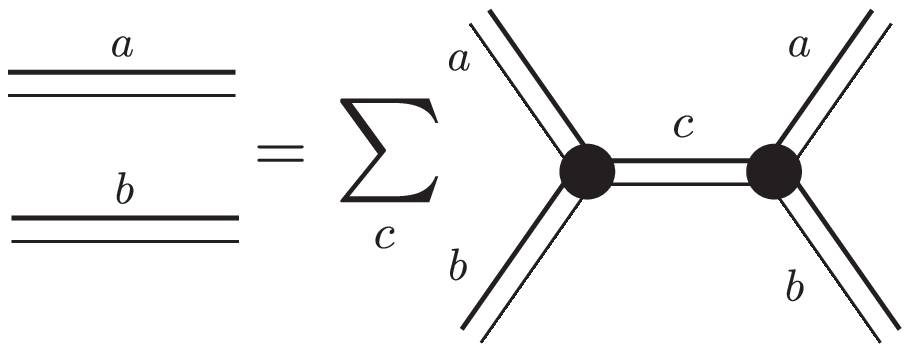}}$\qquad$ \\
 (a)  \\
  \vcenteredhbox{\includegraphics[scale=0.7]{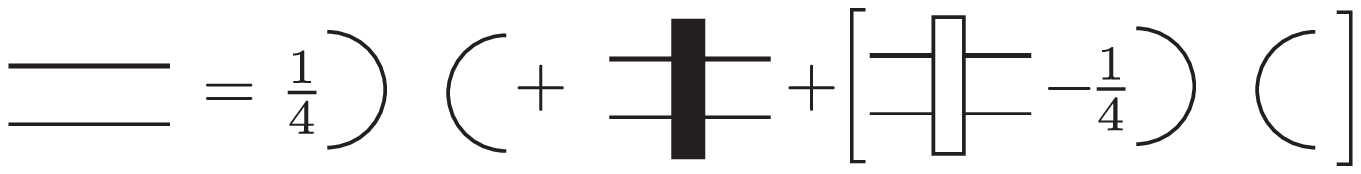}} \\
 (b)  \\
  \vcenteredhbox{\includegraphics[scale=0.7]{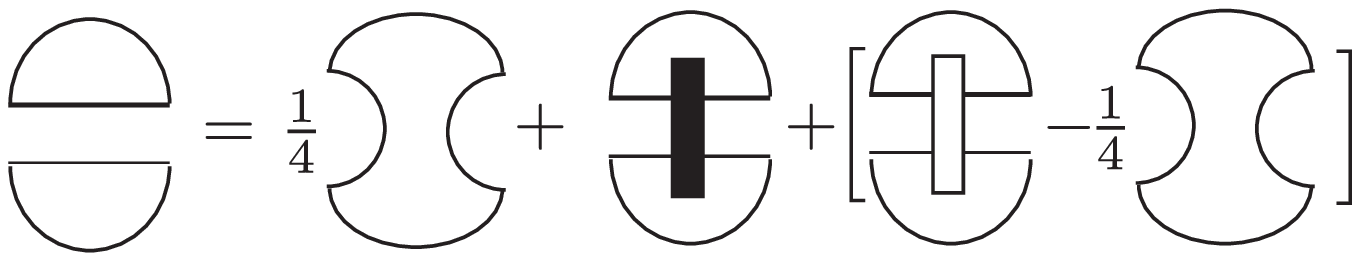}} \\
 (c)
\end{tabular}
 \caption{(a) Generic decomposition of a product of two irreps into a sum of irreps.
              The summation index $c$ runs over all relevant irreps.
          (b) Decomposition of the product of two vector irreps. The white box represents a symmetrizer. 
              Its definition agrees with the antisymmetrizer when removing all minus signs in the definition of the latter.
          (c) Trace of the projector equation (b) leading to the dimension of each irreducible subspace. \label{FIG:ClebshGordonDecomposition}}
\end{figure}

Besides the above equation we need one further property of the mathematical structure realizing the Lorentz symmetry.
A product of any irreps (with representation label $a,b,c,\dots$) is in general reducible and can be decomposed into a sum of irreps
\begin{align}
 g^{(a)}_{\{\mu\}\{\nu\}} \times g^{(b)}_{\{\mu\}\{\nu\}} = \sum_c g^{(c)}_{\{\mu\}\{\nu\}}\,.\label{EQ:ClebshGordonDecomposition}
\end{align}
One can write the above equation in bird track notation like Fig.~\ref{FIG:ClebshGordonDecomposition} (a) shows.
A special case of this decomposition is for example shown in Fig.~\ref{FIG:ClebshGordonDecomposition} (b); the product of two vector representations
\begin{align}
\underbrace{ g_{\mu_1 \nu_1}}_{4} \underbrace{\times g_{\mu_2 \nu_2}}_{4} = &\underbrace{\frac{1}{4}g_{\mu_1 \mu_2}g_{\nu_1 \nu_2}}_{1_S}\nonumber\\
&+\underbrace{\frac{1}{2}(g_{\mu_1 \nu_1}g_{\mu_2 \nu_2}-g_{\mu_1 \nu_2}g_{\mu_2 \nu_1})}_{6_A} + \nonumber\\
&+\underbrace{\frac{1}{2}(g_{\mu_1 \nu_1}g_{\mu_2 \nu_2}+g_{\mu_1 \nu_2}g_{\mu_2 \nu_1})-\frac{1}{4}g_{\mu_1 \mu_2}g_{\nu_1 \nu_2}}_{9_S}\,.
\end{align}
Note that the total dimension of the product space on the l.h.s. has to agree with the sum of all subspaces on the r.h.s of the equation.
Here it is $4 \times 4 = 1+6+9$.
Further each expression on the r.h.s  which represents an independent subspace works as projector on the specific subspace.
This means
\begin{align}
 g^{(a)\,\{\nu\}}_{\{\mu\}} g^{(b)\,\{\rho\}}_{\{\nu\}} &= \delta_{a b} g^{(a)\,\{\rho\}}_{\{\mu\}}\,.
\end{align}
The dimension of a subspace is given by the trace of the projector
\begin{align}
 {\rm Tr}\{ g^{(a)\,\{\nu\}}_{\{\mu\}} \} = g^{(a)\,\{\mu\}}_{\{\mu\}} =d_a \,.
\end{align}
The trace of the projector can be easily evaluated within the bird track notation.
As an example we show this for our example above in Fig.~\ref{FIG:ClebshGordonDecomposition} (c).
The change of the basis which allows to diagonalize the projector of an irreducible subspace is performed with the so called Clebsch-Gordon coefficients $f_{a,b,c}^{{\nu},x}$.
\begin{align}
 f_{(a b c)\,x}^{\{\nu\}}\bigg[g^{(a)\,\{\mu\}}_{\{\nu\}}\times g^{(b)\,\{\mu\}}_{\{\nu\}}\bigg]_{(c)}\left(f^{\{\mu\} y }_{(a b c)}\right)^* = g^{(c)\,y}_{x}\,.
\end{align}
Where the square bracket with index $(c)$ takes only the contribution from the subspace of the irrep $c$ from the argument.
\begin{align}
  \bigg[ X \bigg]_{(c)} =    g^{(c)\,\{\nu\}}_{\{\mu\}} X_{\{\nu\}}^{\{\rho\}} g^{(c)\,\{\sigma\}}_{\{\rho\}}\,. 
\end{align}
The new index $x$ and $y$  label the new diagonal basis indices of irrep $c$. 
The Clebsch-Gordon coefficients can be represented in the bird track notation by a three irrep vertex like depicted in Fig.~\ref{FIG:ClebshGordonDecomposition} (a).

\section{Lorentz basis for Bhabha scattering like amplitudes}\label{SEC:BasisDef}

\begin{figure}
\centering
\begin{tabular}{ccc}
\vcenteredhbox{\includegraphics[scale=0.9]{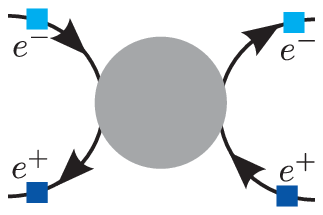}}$\qquad$ &
\vcenteredhbox{\includegraphics[scale=0.7]{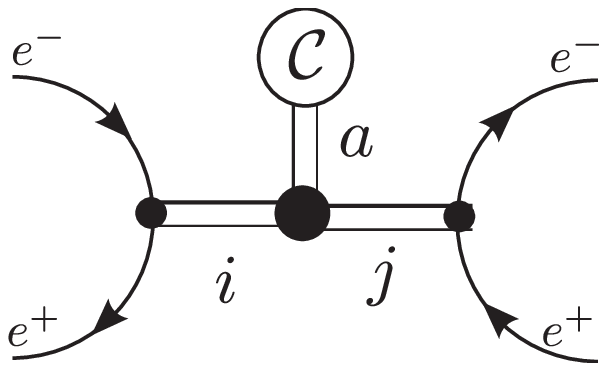}}$\qquad$&
\vcenteredhbox{\includegraphics[scale=0.7]{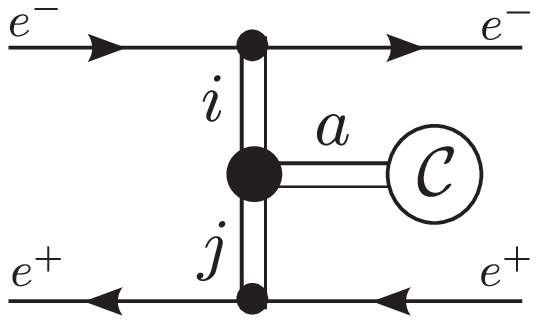}}$\qquad$\\
 (a) & (b) & (c)
\end{tabular}
 \caption{(a) Generic Bhabha scattering amplitude $\mathcal{M}$.\label{FIG:BhabhaScatteringAmps}
          (b) Generic Lorentz structure of a truncated amplitude represented in the s-channel basis.
          (c) Generic Lorentz structure of a truncated amplitude represented in the t-channel basis.}
\end{figure}

In order to construct a general Lorentz basis for Bhabha scattering like amplitudes we show the generic structure of the amplitude $\mathcal{M}$ in Fig.~\ref{FIG:BhabhaScatteringAmps} (a).
Here the gray blobb stands for any internal Lorentz structure.
The four blue rectangles represent the external fermion spinors or, in case of truncated amplitudes (with removed external spinors), the possible respective projector on particle and anti-particle states.
Particles enter the graph on the l.h.s. (initial state)  and exit the graph on the r.h.s  (final state).
The generic notation includes two types of diagrams namely s- and t-channel type diagrams.
In the first case the incoming fermion and anti-fermion line form separately from the outgoing fermion and anti-fermion a spin chain.
In the second case the incoming and outgoing fermion line form separately from the incoming and outgoing anti-fermion a spin chain.

The general Lorentz structure of any truncated amplitude has to respect the given transformation properties. 
That means it transforms as rank four tensor in spinor space.
Because we know that any product of $\gamma$-matrices which appear in a single spin chain can be expressed in terms of a superposition of one of the 16 bilinears,
the most general structure for the amplitude is an expansion in a two bilinear product $\Gamma_i\times\Gamma_j$ with the further condition that there is no open vector index left.
That means the representations originating from the bilinears must combine to form a singlet after the contraction with all available tensors.
In the case where the amplitude is independent of any four momenta this means we have to contract the two bilinear with each other such that they form a singlet.
In case where the amplitude contains additional four vectors like momenta or polarization vectors one can use a linear combination of them in order to contract open indices from the two bilinears.
In order to do this in a systematic way one can decompose the product of two irreps stemming from the bilinears into a sum of irreps.
Thus we arrive at the very general Lorentz structure for an amplitude like depicted in Fig.~\ref{FIG:BhabhaScatteringAmps} (b).
We call the bubble with label $\mathcal{C}$ coefficient tensor and  it stands for a generic Lorentz tensor built up from all available momentum or polarization four vectors.
The generic Lorentz tensor decomposes in dependence of the bilinear irreps $i$ and $j$ into subspaces of the respective irrep $a$ and we have the explicit irrep label and component dependence $\mathcal{C}_{i,j;a}^{\{\mu\}}$.
We choose our bilinears to be in the s-channel setup and we call the resulting basis s-channel basis.
The bird track notation can be cast into the following formula for the truncated amplitude:
\begin{align}
 \mathcal{M} =\sum\limits_{i,j,a}\mathcal{C}_{i,j,a}(\mathcal{M})|i,j;a\rangle_{(s)} \,.
\end{align}
Here we hide any explicit Lorentz tensor structure dependence of the basis elements in the bra-ket notation.
Further we keep all external spinor indices of the basis element $|i,j;a\rangle^{(s)}$ and all potential (four vector) summation indices between basis elements and coefficient tensors implicit.
The  kets and bras are shorthands and have the following definition
\begin{align}
 |i,j;a\rangle_{(s)}^{\{\sigma\}} &= g^{(a)\,\{\sigma\}}_{\{\mu,\nu\}}\Big(e^{+}_{\text{in}}\Big)\Gamma_{i}^{\{\mu\}} \Big(e^{-}_{\text{in}}\Big)
                                                                                   \otimes\Big(e^{-}_{\text{out}}\Big) \Gamma_{j}^{\{\nu\}}\Big(e^{+}_{\text{out}}\Big) \,,\nonumber\\
 {}_{(s)}\langle i,j;a|^{\{\sigma\}} &=\sigma_i\sigma_j g^{(a)\,\{\sigma\}}_{\{\mu,\nu\}} \Big(\overline{e^{-}_{\text{in}}}\Big)\overline{\Gamma}_{i}^{\{\mu\}}\Big(\overline{e^{+}_{\text{in}}}\Big) 
                                                                                  \otimes \Big(\overline{e^{+}_{\text{out}}}\Big)\overline{\Gamma}_{j}^{\{\nu\}}\Big(\overline{e^{-}_{\text{out}}}\Big)\,.
\end{align}
The displayed round brackets containing the incoming and outgoing electron and positron states just define the spin contraction order. 
They are not part of the basis elements.

Alternatively one can also choose a t-channel basis by requiring the bilinears to be in the t-channel setup.
In this case we use the following shorthand for the basis tensor definition
\begin{align}
 |i,j;a\rangle_{(t)}^{\{\sigma\}} &= g^{(a)\,\{\sigma\}}_{\{\mu,\nu\}}\Big(e^{-}_{\text{out}}\Big)\Gamma_{i}^{\{\mu\}} \Big(e^{-}_{\text{in}}\Big)
                                                                                   \otimes\Big(e^{+}_{\text{in}}\Big) \Gamma_{j}^{\{\nu\}}\Big(e^{+}_{\text{out}}\Big) \,,\nonumber\\
 {}_{(t)}\langle i,j;a|^{\{\sigma\}} &=\sigma_i\sigma_j g^{(a)\,\{\sigma\}}_{\{\mu,\nu\}} \Big(\overline{e^{-}_{\text{in}}}\Big)\overline{\Gamma}_{i}^{\{\mu\}}\Big(\overline{e^{-}_{\text{out}}}\Big) 
                                                                                  \otimes \Big(\overline{e^{+}_{\text{out}}}\Big)\overline{\Gamma}_{j}^{\{\nu\}}\Big(\overline{e^{+}_{\text{in}}}\Big)\,.
\end{align}
A pictorial representation of the above ket tensor element can be found in Fig.~\ref{FIG:BhabhaScatteringAmps} (c).

In both choices of the basis $(x)\in \{(s),(t)\}$ the respective tensor coefficient explicitly depends on the open vector indices $\{\sigma\}$ such that the contraction
\begin{align}
 \mathcal{C}_{i,j,a}(\mathcal{M})|i,j;a\rangle_{(x)}&=\mathcal{C}_{i,j,a}^{\{\sigma\}}(\mathcal{M})|i,j;a\rangle_{(x) \{\sigma\}}\,,
\end{align}
has only four open spin indices left.
Here and in the following we use the convention that any missing basis label has to be understood as $(s)$ label.

\begin{figure}
\centering
\begin{tabular}{cc}
\vcenteredhbox{\includegraphics[scale=0.50]{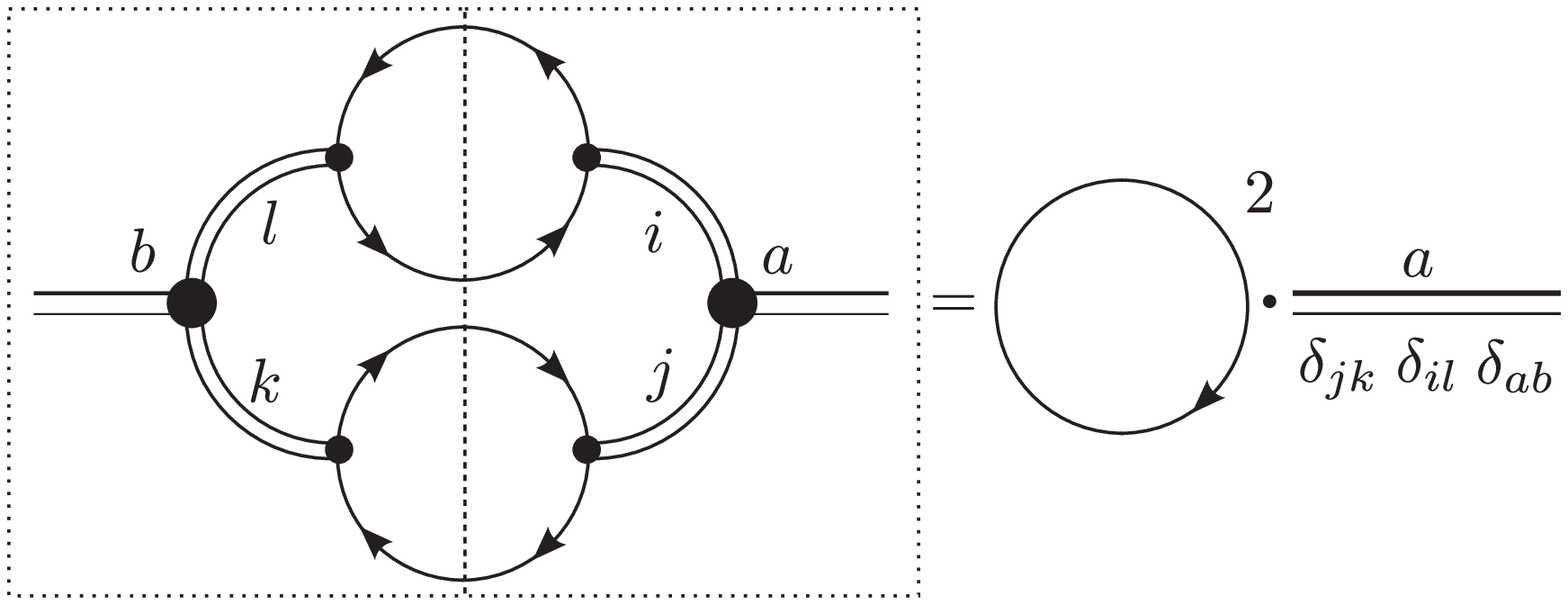}} &  \vcenteredhbox{\includegraphics[scale=0.5]{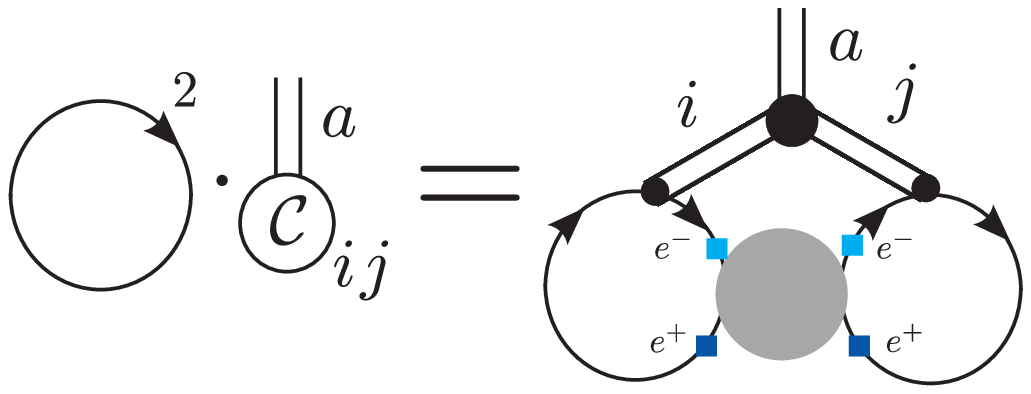}}  \\
  (a) & (b)\\
\end{tabular}
 \caption{(a) Pictorial representation of the orthogonally relation between basis tensors.
          (b) Generic projection on a basis tensor coefficient in the $s$-channel setup.\label{FIG:BasisTensorOrthogonaly}}
\end{figure}

The given basis tensors are block diagonal and fulfill the following orthogonally relation
\begin{align}
   {}_{(x)}^{\{\rho\}}\langle k,l;b|i,j;a\rangle^{\{\sigma\}}_{(x)}=& g^{(b)\,\{\nu_1,\nu_2\}}_{\{\rho\}}g^{(a)\,\{\sigma\}}_{\{\mu_1,\mu_2\}} {\rm tr}\left(\overline{\Gamma}_{k \{\nu_1\}}\Gamma_{i}^{\{\mu_1\}}  \right) {\rm tr}\left(\overline{\Gamma}_{l \{\nu_2\}}\Gamma_{j}^{\{\mu_2\}}  \right)\nonumber\\
    =&d_s^2 g^{(a)\, \{\sigma\}}_{\{\rho\}}  \delta_{ik}\delta_{jl}\delta_{ab}\,.
\end{align}
A pictorial representation of the above equation is shown in Fig.~\ref{FIG:BasisTensorOrthogonaly} (a)
and a more detailed derivation will be given in Sect.~\ref{SEC:AMP2}.
With the given relation it is straight forward to extract the coefficient tensors $\mathcal{C}_{i,j,a}$ from any given amplitude
via the projection
\begin{align}
 \mathcal{C}_{i,j,a}(\mathcal{M})=d_s^{-2}{}_{(x)}\langle i,j;a|\mathcal{M}\,.
\end{align}
As an example of how the suppressed spin index contractions are to be carried out we depict it in case of the $s$-channel setup in Fig.~\ref{FIG:BasisTensorOrthogonaly} (b). 

Due to our convention of absorbing the signs $\sigma_i$ and $\sigma_j$ into the bras we have a non-trivial sign dependent relation between the 
coefficient tensor  $\mathcal{C}_{ija}(\mathcal{M})$ of the amplitude and its co-amplitude analog $\mathcal{C}^{\star}_{i,j,a}$
which appears in the decomposition of the co-amplitude
\begin{align}
\mathcal{M}^{*}= \sum\limits_{i,j,a}\mathcal{C}^{\star}_{i,j,a}(\mathcal{M})\langle i,j;a|\,.
\end{align}
The co-amplitude coefficient tensor can be obtained via the projection
\begin{align}
 \mathcal{C}^{\star}_{i,j,a}(\mathcal{M}) = \mathcal{M}^{*}|i,j;a\rangle\,.
\end{align}
The tensors are directly related to each other via
\begin{align}
 \mathcal{C}^{\star}_{i,j,a}(\mathcal{M})=\sigma_i \sigma_j \mathcal{C}^{*}_{i,j,a}(\mathcal{M})\,.
\end{align}
During the complex conjugation one has the replacement $\tilde{\varepsilon}^*=-\tilde{\varepsilon}$.

\section{Explicit basis tensors}
Up to now we just introduced a generic framework for the decomposition of amplitudes dictated by the algebra of the underlying symmetry.
That means we did not consider any Lorentz group specific properties in order to obtain an explicit representation of the basis elements in terms of bilinears and combinations of the metric tensors $g_{\mu \nu}$.
In order to do so we list in Tab.~\ref{TAB:DecompositionListings} all possible combinations of bilinears (the definition of representation labels $\mathcal{R}_i$ in dependence of the index $i$ are given in Tab.~\ref{TAB:RepLabelListing}).
Further we give a link to the required decomposition into irreps in case it is not a trivial one.
Because the explicit expressions for the decompositions in terms of the metric tensors become quite lengthy, we provide them only in terms of bird tracks.
All decompositions can be found in Ref.~\cite{Cvitanovic:2008zz}.
\begin{table}
\centering
\begin{tabular}{c|c|cc|c}
$\mathcal{R}_i$   & $\mathcal{R}_j$   &  $\mathcal{R}_a$ &  & $d$\\
\hline
$1^x$ & $1^y$ & $1$& & $4\times1$\\
$4^x$ & $1^y$ & $4$& & $4\times4$\\
$1^x$ & $4^y$ & $4$& & $4\times4$\\
$1^x$ & $6$   & $6$& & $2\times6$\\
$6$   & $1^y$ & $6$& & $2\times6$\\
$4^x$ & $4^y$ & $1+6+9$  & Fig.~\ref{FIG:ClebshGordonDecomposition} (a) & $4\times16$\\
$4^x$ & $6$   & $4_a+4_b+16$&Fig.~\ref{FIG:ClebshGordonDecomposition2} (a) & $2\times24$\\
$6$   & $4^y$ & $4_a+4_b+16$&Fig.~\ref{FIG:ClebshGordonDecomposition2} (a) & $2\times24$\\
$6$   & $6$   & $1_a+1_b+6+9_a+9_b+10$&Fig.~\ref{FIG:ClebshGordonDecomposition2} (b) & $1\times36$\\
\end{tabular}
\caption{Decomposition of bilinear products into irreps. The numbers reflect the dimension of the irrep. $x$ and $y$ can be either $-1$ or $+1$.
In case of $-1/+1$ the corresponding bilinear does/does not contain $\gamma^5$. 
$6$ is the adjoint representation emitted from a spin chain via $\Gamma_{\underline{2}}$.
In case we have two representations with equivalet dimension appearing in the decomposition we distinguish them with additional labels $a$ and $b$.
In the last column we state the dimensions of the subspace contained in each decomposition type.\label{TAB:DecompositionListings}
}
\end{table}
\begin{table}
\centering
\begin{tabular}{c|c|c}
$i$   & $\mathcal{R}_i$ & $d_i$\\
\hline
$\underline{0}$ & $1^+$&  $1$\\
$\underline{1}$ & $4^+$&  $4$\\
$\underline{2}$ & $6$&  $6$\\
$\underline{3}$ & $4^-$ & $4$\\
$\underline{4}$ & $1^-$ & $1$  
\end{tabular}
\caption{Definition of the used representation labels $\mathcal{R}_i$ and respective dimension $d_i$ in depedence of the index $i$.\label{TAB:RepLabelListing}
}
\end{table}

In the last column of Tab.~\ref{TAB:DecompositionListings} we also provide the total dimension of the subspaces contained in each decomposition type.
From the table we see that the total dimension of the Lorentz basis is $256$ (in four space-time dimensions).
However, due to block diagonal structure of the basis the evaluation of $256 \times 256 = 65536$ element product space of two amplitudes
reduces to the evaluation of a
\begin{align}
 4\cdot 1 + 4\cdot4^2 + 4\cdot4^2 + 2\cdot6^2 + 2\cdot6^2 + 4\cdot(1 + 6^2 + 9^2)& \nonumber\\
 + 4\cdot(4^2 + 4^2 + 16^2) + 1\cdot(1 + 1 + 6^2 + 9^2 + 9^2 + 10^2)&=2200\,
\end{align}
element product space.
One can in principle go further and diagonalize all sub-blocks in terms of orthogonal basis elements to reduce this number to $256$.
However, the diagonalization of all sub-blocks goes beyond this paper, and we will only work with components written  in terms of the four-dimensional vector representation.
\begin{figure}
\centering
\begin{tabular}{cc}
\vcenteredhbox{\includegraphics[scale=0.5]{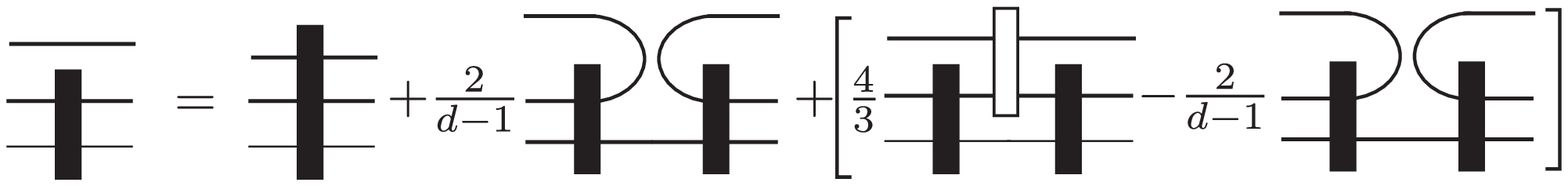}}$\qquad$ \\
 (a)  \\
  \vcenteredhbox{\includegraphics[scale=1.0]{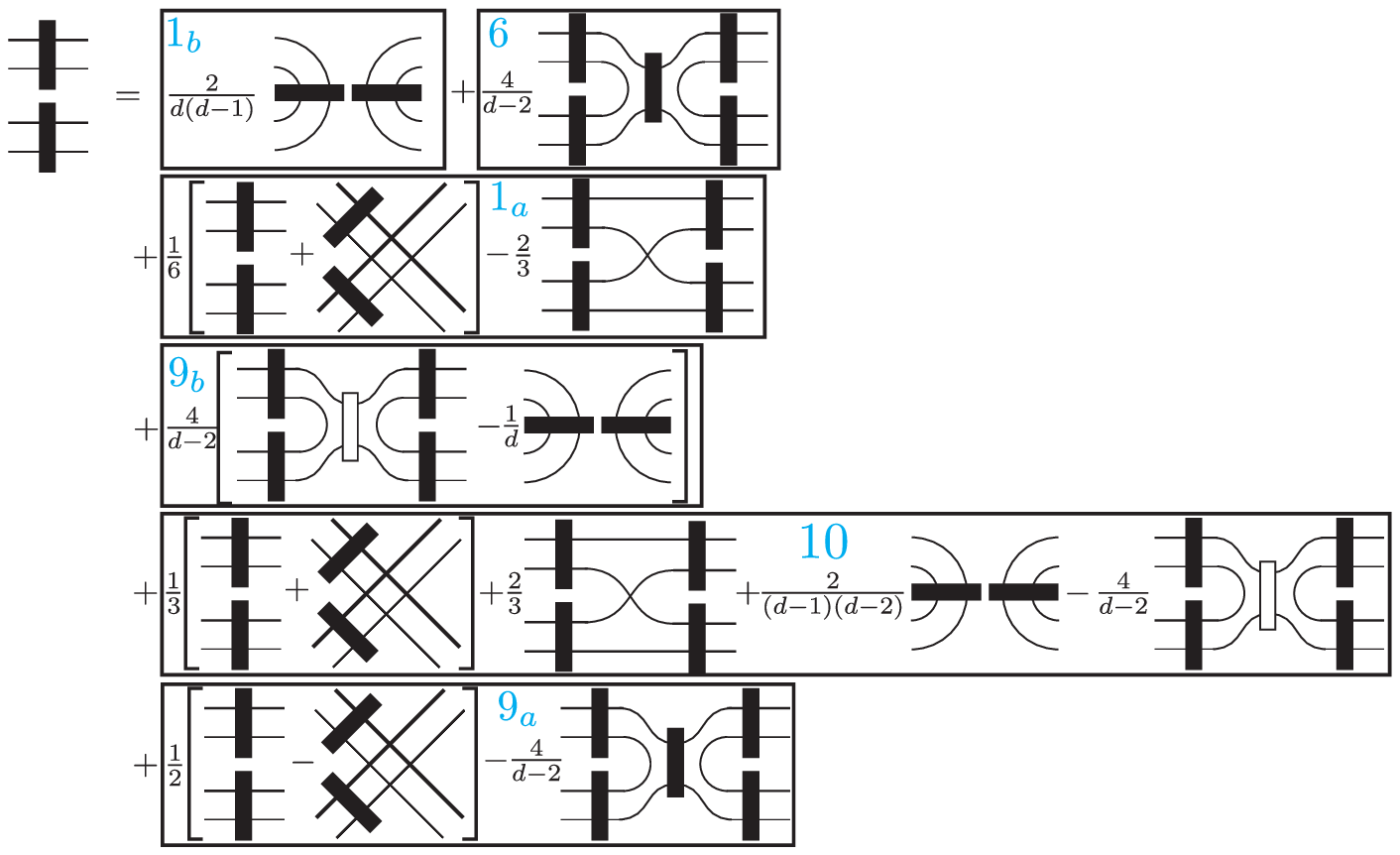}} \\
 (b) \\
  \vcenteredhbox{\includegraphics[scale=0.5]{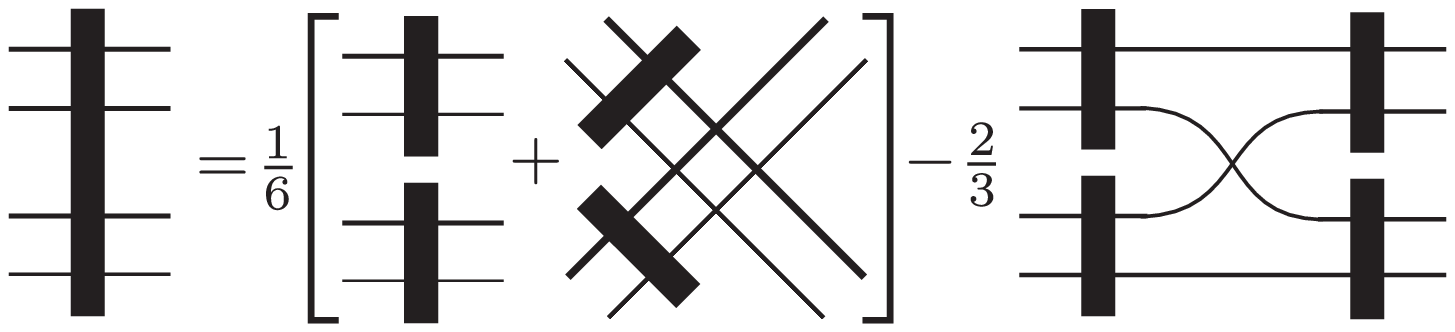}} \\
 (c)
\end{tabular}
 \caption{(a) Decomposition of $4\times6$ product space.
          (b) Decomposition of $6\times6$ product space. The light blue numbers indicate the dimension of the irrep in case the vector representation has dimension $d=4$.
          (c) Alternative representation of irrep $1_a$.\label{FIG:ClebshGordonDecomposition2}}
\end{figure}

We further state the parity for each of our basis tensors elements
\begin{align}
 P_{\varepsilon}(|i,j,a\rangle)= (-1)^{\delta_{\underline{3}i}+\delta_{\underline{4}i}+\delta_{\underline{3}j}+\delta_{\underline{4}j}}\,.
\end{align}
Every Dirac bilinear that contains a $\gamma^5$ automatically contains an $\varepsilon$-tensor.
Further the representation index $a$ has no influence on the parity because each basis element 
contains the projector onto the representation $a$ and thus always contains an even number of $\varepsilon$-tensors.
As an example one can have a look at the decomposition of $6\times 6$ irrep product in Fig.~\ref{FIG:ClebshGordonDecomposition2} (b).
There we clearly see, using the relation of Fig.~\ref{FIG:ClebshGordonDecomposition2} (c) and Fig.~\ref{FIG:UseFulRelations} (a), 
that the representation $1_a$ is indeed the projector on the pseudoscalar subspace $1^-$,
however the projector obviously contains two $\varepsilon$-tensors.

\section{General Fierz transformations}
It can be shown via the insertion of two completeness relation between legs of different spin chains and the algebraic reduction of the appearing (loop) tensor structure,
involving a single trace over four bilinears, that any $(s)$ basis tensor can be represented via a linear combination of $(t)$ channel basis tensors and vice versa.
The resulting transformations are called Fierz transformation and were introduce in Ref.~\cite{Fierz:1937}.
Explicit Fierz coefficients for the scalar and pseudoscalar case can be found in Ref.~\cite{Nieves:2003in} and~\cite{Nishi:2004st}.
Fierz transformation for all higher tensor bilinears are calculated in Ref.~\cite{Liao:2012uj}.
We can calculate the Fierz coefficients using the presented basis.
The main advantage of the given basis is that we automatically obtain the coefficients in a block-diagonal form,
where all entries with inequivalent parity and dimension of representation $a$ and $b$ vanishes.

In the case of $\mathcal{R}_a=1^+$  we obtain:
\begin{align}
 \left(\begin{array}{c}|\underline{0},\underline{0};\underline{0}\rangle_{(t)}^{\prime} \\ 
 |\underline{1},\underline{1};\underline{0}\rangle_{(t)}^{\prime} \\ 
 |\underline{2},\underline{2};\underline{0}\rangle_{(t)}^{\prime} \\ 
 |\underline{3},\underline{3};\underline{0}\rangle_{(t)}^{\prime} \\ 
 |\underline{4},\underline{4};\underline{0}\rangle_{(t)}^{\prime}\end{array}\right)= 
 \left(\begin{array}{ccccc}
 +\frac{1}{4} & +\frac{1}{4} & +\frac{1}{4} & -\frac{1}{4}  & +\frac{1}{4}\\ 
 +1 & -\frac{1}{2} & 0 & -\frac{1}{2}  & -1 \\ 
 +\frac{3}{2} & 0 & -\frac{1}{2} & 0  & +\frac{3}{2}\\
 -1 & -\frac{1}{2} & 0 & -\frac{1}{2}  & 1 \\ 
 +\frac{1}{4} & -\frac{1}{4} & +\frac{1}{4} & +\frac{1}{4}  & +\frac{1}{4}\\\end{array}\right)
 \left(\begin{array}{c}|\underline{0},\underline{0};\underline{0}\rangle_{(s)}^{\prime} \\ 
 |\underline{1},\underline{1};\underline{0}\rangle_{(s)}^{\prime} \\ 
 |\underline{2},\underline{2};\underline{0}\rangle_{(s)}^{\prime} \\ 
 |\underline{3},\underline{3};\underline{0}\rangle_{(s)}^{\prime} \\ 
 |\underline{4},\underline{4};\underline{0}\rangle_{(s)}^{\prime}\end{array}\right)\,,
\end{align}
where we introduce the primed basis via the absorbtion of the singlet coefficient tensor $g^{\mu \nu}$ into the basis tensors
\begin{align}
 |i,i;\underline{0}\rangle^{\prime} = g^{(i)}_{\{\mu\}}|i,i;\underline{0}\rangle^{\{\mu\}}\,.
\end{align}
This changes the orthogonality relation to
\begin{align}
 {}^{\prime}_{(x)}\langle k,l;\underline{0}|i,j;\underline{0}\rangle^{\prime}_{(x)} = d_s^2d_i\delta_{ij}\delta_{kl}\delta_{ik}\,.
\end{align}
The coefficients in the matrix above can be obtained via:
\begin{align}
 |i,i;\underline{0}\rangle_{(t)}^{\prime}=\sum\limits_{j=\underline{0}}^{\underline{4}}|j,j;\underline{0}\rangle_{(s)}^{\prime}{}_{(s)}^{\prime}\langle j,j;\underline{0}|i,i;\underline{0}\rangle_{(t)}^{\prime}\,,
\end{align}
which is just the completeness relation of our s-channel tensor basis applied to the singlet subspace.

We can generalize the above equation to the case where $a\neq0$ that means where we have in general:
\begin{align}
 |i,j;a\rangle_{(t)}=\sum\limits_{k,l=\underline{0}}^{\underline{4}}\sum\limits_{b}|k,l;b\rangle_{(s)}{}_{(s)}\langle k,l;b|i,j;a\rangle_{(t)}\,.
\end{align}
The product of the s- and t-basis elements represent the generalization of the Fierz coefficient to the general $a$ (non-singlet) case.
\begin{figure}
\centering
\begin{tabular}{c}
 \includegraphics[scale=0.6]{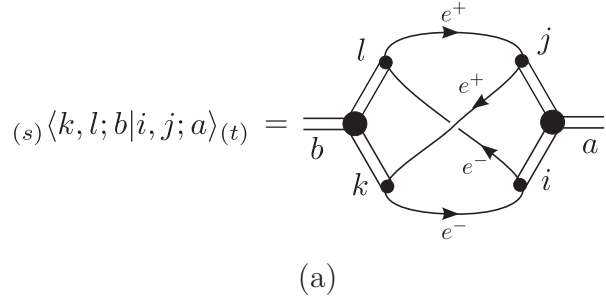}\\
 (a)
\end{tabular}
\caption{(a) Bird track notation for the general Fierz coefficients. \label{FIG:DefFierzCoeff}}
\end{figure}
The tensor coefficients appearing in the basis change are defined in bird track notation in Fig.~\ref{FIG:DefFierzCoeff}.
It is clear that due to the properties of the Lorentz symmetry the coefficients can be non-vanishing only when the dimension of the representation $a$ and $b$ do agree.
This serves as welcome check for our explicit calculation, where we calculate the Fierz coefficients to all possible combinations of s- and t-basis elements within FORM~\cite{Kuipers:2012rf}.
Here one has to mention that in fact the projection leads to non-trivially vanishing expressions in the case where $a=4$ and $b=16$ and $a=6$ and $b=9$ (and vice versa).
That means one has to apply non-trivial relations which hold due to symmetry in order to make the zero explicit.
For the reader's convenience we provide more details on the treatment of these expressions in Appendix~\ref{AppendixICSI}.

%
As another example the pseudoscalar subspace ($a=\underline{4}, \mathcal{R}_{\underline{4}}=1^-$) Fierz matrix can be displayed in terms of a scalar matrix after the absorption of all metric tensors and Levi-Civita tensors with tilde into the primed basis elements.
Here we explicitly define 
\begin{align}
 |\underline{2},\underline{2};\underline{4}\rangle^{\prime}= \sqrt{6}\tilde{\epsilon}^{\mu_1\mu_2\mu_3\mu_4} |\underline{2},\underline{2};1_a\rangle_{\mu_1\mu_2\mu_3\mu_4}\,.
\end{align}
The Fierz matrix reads on the subspace
\begin{align}
 \left(\begin{array}{c}
 |\underline{0},\underline{4};\underline{4}\rangle_{(t)}^{\prime} \\ 
 |\underline{1},\underline{3};\underline{4}\rangle_{(t)}^{\prime} \\ 
 |\underline{2},\underline{2};\underline{4}\rangle_{(t)}^{\prime} \\ 
 |\underline{3},\underline{1};\underline{4}\rangle_{(t)}^{\prime} \\ 
 |\underline{4},\underline{0};\underline{4}\rangle_{(t)}^{\prime}\end{array}\right)= 
 \left(\begin{array}{ccccc}
 +\frac{1}{4} & +\frac{1}{4} & -\frac{1}{4} & -\frac{1}{4}  & +\frac{1}{4}\\ 
 +1 & -\frac{1}{2} & 0 & -\frac{1}{2}  & -1 \\ 
 -\frac{3}{2} & 0 & -\frac{1}{2} & 0  & -\frac{3}{2}\\
 -1 & -\frac{1}{2} & 0 & -\frac{1}{2}  & +1 \\ 
 +\frac{1}{4} & -\frac{1}{4} & -\frac{1}{4} & +\frac{1}{4}  & +\frac{1}{4}\\\end{array}\right)
 \left(\begin{array}{c}
 |\underline{0},\underline{4};\underline{4}\rangle_{(s)}^{\prime} \\ 
 |\underline{1},\underline{3};\underline{4}\rangle_{(s)}^{\prime} \\ 
 |\underline{2},\underline{2};\underline{4}\rangle_{(s)}^{\prime} \\
 |\underline{3},\underline{1};\underline{4}\rangle_{(s)}^{\prime} \\ 
 |\underline{4},\underline{0};\underline{4}\rangle_{(s)}^{\prime}\end{array}\right)\,.
\end{align}
We have in the chosen normalization
\begin{align}
 {}^{\prime}\langle k,l;\underline{4}|i,j;\underline{4}\rangle^{\prime} = d_s^2d_i\delta_{ik}\delta_{lj}\,.
\end{align}
\begin{table}
\centering
\begin{tabular}{c|c}
 $\mathcal{R}_a$ & $|i,j;a\rangle$\\
\hline
 $1^+$&  $|\underline{0},\underline{0};1\rangle,
          |\underline{1},\underline{1};1\rangle,
          |\underline{2},\underline{2};1_b\rangle,
          |\underline{3},\underline{3};1\rangle,
          |\underline{4},\underline{4};1\rangle$\\
 $1^-$&  $|\underline{0},\underline{4};1\rangle,
          |\underline{1},\underline{3};1\rangle,
          |\underline{2},\underline{2};1_a\rangle,
          |\underline{3},\underline{1};1\rangle,
          |\underline{4},\underline{0};1\rangle$\\
 $4^+$&  $|\underline{0},\underline{1};4\rangle, 
          |\underline{4},\underline{3};4\rangle, 
          |\underline{1},\underline{2};4_b\rangle, 
          |\underline{3},\underline{2};4_a\rangle, 
          |\underline{2},\underline{3};4_a\rangle, 
          |\underline{2},\underline{1};4_b\rangle, 
          |\underline{3},\underline{4};4\rangle,
          |\underline{1},\underline{0};4\rangle$\\
 $4^-$&  $|\underline{0},\underline{3};4\rangle, 
          |\underline{4},\underline{1};4\rangle, 
          |\underline{3},\underline{2};4_b\rangle, 
          |\underline{1},\underline{2};4_a\rangle, 
          |\underline{2},\underline{1};4_a\rangle, 
          |\underline{2},\underline{3};4_b\rangle, 
          |\underline{1},\underline{4};4\rangle,
          |\underline{3},\underline{0};4\rangle$\\
 $6$&  $|\underline{0},\underline{2};6\rangle, 
          |\underline{4},\underline{2};6\rangle, 
          |\underline{1},\underline{1};6\rangle, 
          |\underline{1},\underline{4};6\rangle, 
          |\underline{2},\underline{2};6\rangle, 
          |\underline{4},\underline{1};6\rangle, 
          |\underline{3},\underline{3};6\rangle,
          |\underline{2},\underline{4};6\rangle,
          |\underline{2},\underline{0};6\rangle$\\
 $9^+$&  $|\underline{0},\underline{0};9\rangle, 
          |\underline{2},\underline{2};9_b\rangle,
          |\underline{3},\underline{3};9\rangle$\\
 $9^-$&  $|\underline{0},\underline{3};9\rangle, 
          |\underline{2},\underline{2};9_a\rangle,
          |\underline{3},\underline{0};9\rangle$\\
 $10$&   $|\underline{2},\underline{2};10\rangle$\\
\end{tabular}
\caption{Listing of all independent Fierz sub-blocks in dependence of the irrep $\mathcal{R}_a$ and the involved basis elements $|i,j;a\rangle$ which transform into each other under basis change $(s) \leftrightarrow (t)$ \label{TAB:FierzBlockListing}.
}
\end{table}

In Table~\ref{TAB:FierzBlockListing} we list all Fierz sub matrices including the involved basis elements transforming into each other during the $(s) \leftrightarrow (t)$ basis change.
In general only tensor basis elements with equivalent irrep $a$ dimension $d_a$ do mix under Fierz transformation with each other.
Besides the dimension there is one further selection rule, which is based on the number parity of a given basis element.
Any decomposition of a generic Lorentz tensor $T$ into a sum over a product of tensors
has a well defined parity.
That means every term appearing in the decomposition must have the same parity which is given by $ P_{\varepsilon}(T)$.
A Fierz transformation $\mathcal{F}$ obeys the above rule and we have
$P_{\varepsilon}(\mathcal{F}(|i,j,a\rangle))=P_{\varepsilon}(|i,j,a\rangle)$.

This immediately explains why the $a=\underline{0}$ (alias $1^+$) and $a=\underline{4}$  (alias $1^-$) case 
decouple from each other although the dimension of both representation is given by one.
In the $1^-$ case we further see that a state with odd parity can decompose into a state of even parity in case the coefficient tensor (which we absorbed above into the primed ket)
has odd parity.
However, this can only happen if the number and symmetry of independent Lorentz indices of a given basis tensor is appropriate to form a contraction with an $\varepsilon$-tensor so that the original tensor structure is retained. 

The four-dimensional subspace splits into two separate sub-blocks $4^+$ and $4^-$, too.
We find eight basis tensors in each of them.

In the six-dimensional subspace we have nine states. 
Because here the index setup leaves options to have transitions from all odd to all even parity basis tensors we do not have a
separation into even and odd sub-blocks.

The nine-dimensional subspace splits again into even and odd subspaces where we have three basis tensors in each of them.

The ten-dimensional subspace is only populated through the $\Gamma_{\underline{2}}\times\Gamma_{\underline{2}}$ decomposition.
Thus there is only one basis tensor.

\section{Products of Bhabha scattering amplitudes}\label{SEC:AMP2}

\begin{figure}
\centering
\begin{tabular}{cc}
\vcenteredhbox{\includegraphics[scale=0.9]{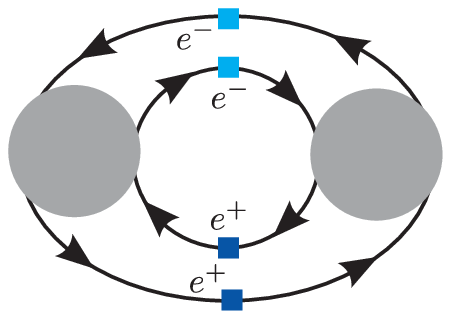}} &  \vcenteredhbox{\includegraphics[scale=0.5]{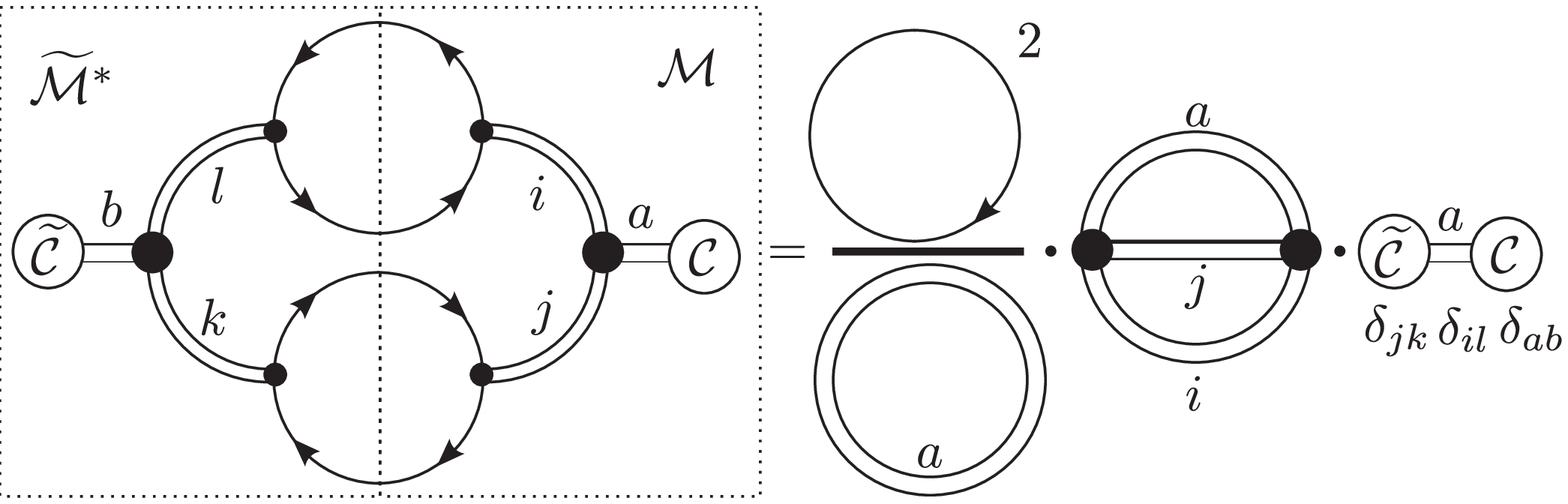}}  \\
  (a) & (b)
\end{tabular}
 \caption{(a) Generic Bhabha scattering amplitude product $ \mathcal{M}_1^{*}\mathcal{M}_2$.
          (b) Evaluation of the amplitude product within the irreducible Lorentz tensor basis.\label{FIG:BhabhaScatteringAmpSquare} }
\end{figure}

In the following we demonstrate how to evaluate squares of amplitude within the introduced basis.
In Fig.~\ref{FIG:BhabhaScatteringAmpSquare} (a) we show the contraction prescription of the external legs that has to be carried out in order to arrive at the product of two amplitudes in general.
The gray blobs indicates a generic Lorentz amplitude.
The small (light blue)/blue rectangles indicate projectors on a specific spin and or (anti-fermion)/fermion state.
They follow for example from the completeness relation of external particle and anti-particle spinors.

In Fig.~\ref{FIG:BhabhaScatteringAmpSquare} (b) we evaluate the product of two (in general different) amplitudes with tensor coefficients $\mathcal{C}$ and $\tilde{\mathcal{C}}$.
In case both amplitudes coincide we thus calculate the amplitude square.
Here we implicitly assume that we absorbed any appearing projectors of external legs in one of the two or both amplitudes.

To arrive at the right hand side of the equation we use the orthogonally relation of the bilinears and the fact that there are only diagonal rank two tensors in representation space.
In the language of bird tracks this means different representation can only be connected via a vertex, which in turn can be understood as a Clebsch-Gordon decomposition.
All tensors with only two indices vanish in case the indices belong to a different representation.
The bird track equation translates into:
\begin{align}
\widetilde{\mathcal{M}}^{*} \cdot\mathcal{M}=\sum\limits_{i,j,a}\sum\limits_{l,k,b}\widetilde{\mathcal{C}}^{\star}_{klb} \langle l,k;b|i,j;a\rangle \mathcal{C}_{ija}  =\sum\limits_{i,j,a} \frac{d_s^2}{d_a} \mathcal{J}^{(3)}_{ija} \left(\widetilde{\mathcal{C}}^{\star}_{jia}\cdot\mathcal{C}_{ija}\right) \,.
\end{align}
One can evaluate the group specific numbers $\mathcal{J}^{(3)}_{ija}$, which are related to the Wigner-$3j$ symbols, by taking the trace over the projector on the irrep $a$ appearing in the decomposition $i\otimes j\rightarrow a$ .
In our normalization we simply have
\begin{align}
 \mathcal{J}^{(3)}_{ija}=d_a\,,
\end{align}
because the trace over the projector onto the irrep $a$ is equal to its dimension.
Further we conclude that the basis tensors are indeed orthogonal 
\begin{align}
 \langle l,k;b|i,j;a\rangle = g^{(a)} \delta_{i l}\delta_{j k}\delta_{a b} \,,
\end{align}
and we can square any amplitude once we know all coefficient tensors $\mathcal{C}_{ija}$.
This has the advantage that the obtained coefficient tensor is free of any spin structure and contains the full amplitude information.
It is thus not necessary to perform an explicit amplitude square calculation in order to be able to use the trace technology for general spin structures.
With the chosen normalization convention of the Dirac bilinears all the coefficient tensors are without explicit factors of imaginary units 
(using the tensor $\tilde{\varepsilon}$ instead of $\varepsilon$) if the original amplitude is real.
This is the case if it has neither an imaginary part due to above threshold kinematics inside a loop diagram nor any complex couplings.

\section{Example: Tree-level Bhabha scattering }
\begin{figure}
\centering
\begin{tabular}{c}
 \includegraphics[scale=0.5]{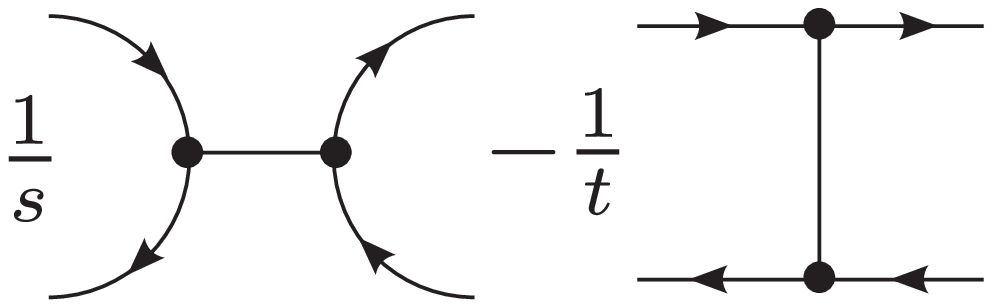}\\
 (a)\\
 \includegraphics[scale=0.5]{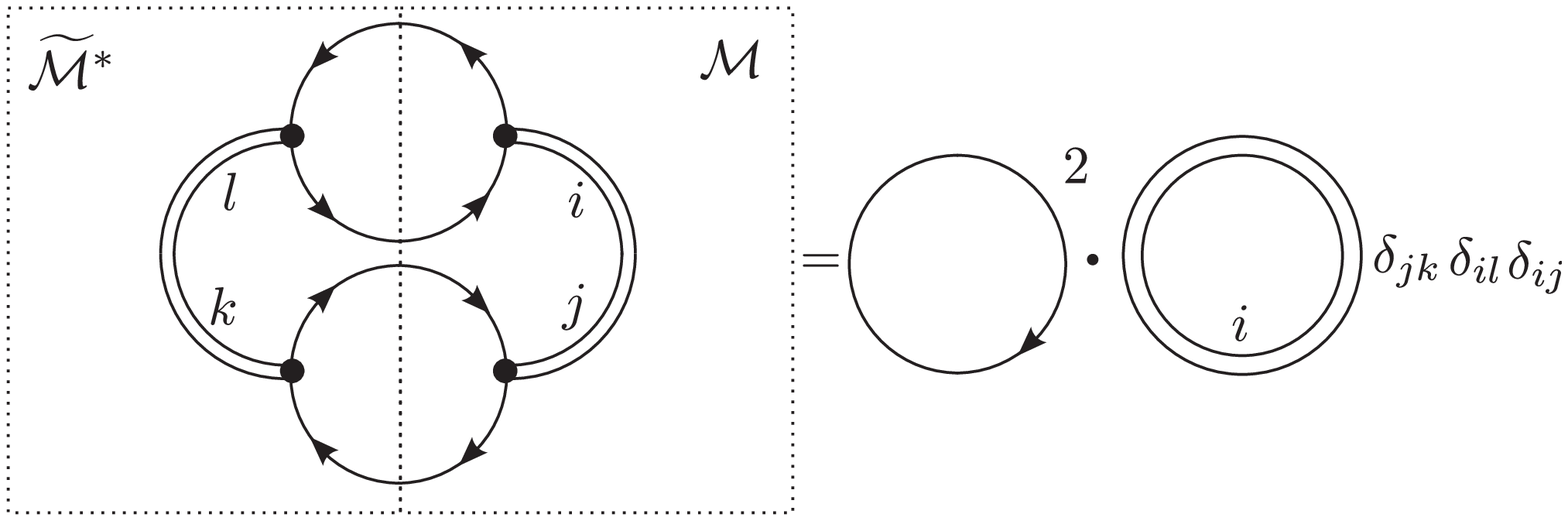} \\
 (b)
\end{tabular}
\caption{(a) Lorentz structure of the Born amplitude.
         (b) Evaluation of the squared amplitude within the irreducible Lorentz tensor basis in the special singlet case ($a=\underline{0}$). \label{FIG:BhabhaScatteringAmpsSpecial}}
\end{figure}

When calculating higher order corrections to the unpolarized Bhabha scattering cross section a generic contribution is given by the product of the Born amplitude with any higher order amplitude to this order.
In this product one can choose to absorb the external on-shell fermion and anti-fermion state projectors into the higher order amplitudes.
One is then left with the tree-level amplitude like shown in Fig.~\ref{FIG:BhabhaScatteringAmpsSpecial} (a).
Here we have an example for the special case where the amplitude does not depend on any external momenta.
We have (up to an overall factor) using the Mandelstam variables $s=(p_1+p_2)^2=(p_3+p_4)^2$ and $t=(p_1-p_2)^2=(p_3-p_4)^2$ 
\begin{align}
 \widetilde{\mathcal{M}}^{(\text{tree})}\sim \frac{1}{s}(\gamma^\mu)^{\alpha_2}_{\,\alpha_1}(\gamma_{\mu})^{\alpha_3}_{\,\alpha_4} - \frac{1}{t}(\gamma^\mu)^{\alpha_3}_{\,\alpha_1}(\gamma_{\mu})^{\alpha_2}_{\,\alpha_4}\,.
\end{align}
There are only contributions from the amplitude for $\mathcal{R}_a=1^+$ ($a=\underline{0}$),
meaning the decomposition of the two irreps has to yield a singlet.
This immediately restricts the selection of $i$ and $j$ to the case $i=j$.
The amplitudes read in our basis:
\begin{align}
 \widetilde{\mathcal{M}}^{(\text{tree-}s)}&=+\frac{1}{s}g^{\mu_1\mu_3}|\underline{1},\underline{1};\underline{0}\rangle_{\mu_1 \mu_3},\nonumber\\
 \widetilde{\mathcal{M}}^{(\text{tree-}t)}&=-\frac{1}{t}|\underline{0},\underline{0};\underline{0}\rangle
                                            +\frac{1}{2t}g^{\mu_1\mu_3}|\underline{1},\underline{1};\underline{0}\rangle_{\mu_1 \mu_3}
                                            +\frac{1}{2t}g^{\mu_1\mu_3}|\underline{3},\underline{3};\underline{0}\rangle_{\mu_1 \mu_3}
                                            +\frac{1}{t}|\underline{4},\underline{4};\underline{0}\rangle\,.
\end{align}
The above result has an important consequence.
Any amplitude square produced with the above unpolarized tree-level amplitude will have contributions from the sub-blocks corresponding to $|\underline{0},\underline{0};\underline{0}\rangle,|\underline{1},\underline{1};\underline{0}\rangle,|\underline{3},\underline{3};\underline{0}\rangle,|\underline{4},\underline{4};\underline{0}\rangle$, only.
That means in this case we only need to evaluate four instead of $256$ coefficients in the respective co-amplitude.

We further want to note that it is convenient to absorb the stated metric tensors appearing in the tensor coefficients into the kets.
This is equivalent to switching to the primed basis introduced in Sect.~4. 
In the primed basis we thus have
\begin{align}
 \widetilde{\mathcal{M}}^{\text{tree}} = \sum_{i} \tilde{c}_{ii\underline{0}}|i,i;\underline{0}\rangle^{\prime}\,,
\end{align}
with the following non-vanishing scalar coefficients ($\tilde{c}_{iia}=\tilde{c}^{(\text{tree-}s)}_{iia}+\tilde{c}^{(\text{tree-}t)}_{iia}$):
\begin{align}
\tilde{c}^{(\text{tree-}s)}_{\underline{110}}&=+\frac{1}{s}\,,\\
\tilde{c}^{(\text{tree-}t)}_{\underline{000}}&=-\frac{1}{t}\,,&  
\tilde{c}^{(\text{tree-}t)}_{\underline{110}}&=+\frac{1}{2t}\,,&   
\tilde{c}^{(\text{tree-}t)}_{\underline{330}}&=+\frac{1}{2t}&  
\tilde{c}^{(\text{tree-}t)}_{\underline{440}}&=+\frac{1}{t}\,.
\end{align}
In this case we can construct our amplitude product following Fig.~\ref{FIG:BhabhaScatteringAmpsSpecial} (b).
That means:
\begin{align}
 {}^{\prime}\langle k,l;\underline{0}|i,j;\underline{0}\rangle^{\prime} = d_s^2d_i\delta_{ij}\delta_{kl}\delta_{ik}\,.
\end{align}

For the calculation of the spin averaged born matrix element square $|\mathcal{M}|^2$ we further need to calculate the decomposition of the amplitude including the external particle anti-particle projectors
which follow from the spinor completeness relations
\begin{align}
  \mathcal{M}^{\text{tree}} =&  \frac{1}{s}[(\slashed p_2 -m)\gamma^\mu(\slashed p_1 +m)]^{\alpha_2}_{\,\alpha_1}[(\slashed p_3 +m)\gamma_{\mu}(\slashed p_4 -m)]^{\alpha_3}_{\,\alpha_4} \nonumber\\
   &- \frac{1}{t}[(\slashed p_3 +m)\gamma^\mu(\slashed p_1 +m)]^{\alpha_3}_{\,\alpha_1}[(\slashed p_2 -m)\gamma_{\mu}(\slashed p_4 -m)]^{\alpha_2}_{\,\alpha_4}\,.
\end{align}
With it the amplitude square can be written as
\begin{align}
 |\mathcal{M}^{\text{tree}}|^2 = \widetilde{\mathcal{M}}^{\text{tree} *} \cdot \mathcal{M}^{\text{tree}}= d_s^2\sum\limits_{i} d_i \tilde{c}^{(\text{tree})}_{ii 0} c^{(\text{tree})}_{ii 0}\,,
\end{align}
where the dot indicates the contraction over all four spin indices.
Here we used that for the relevant basis tensors we indeed have $\tilde{c}^{(\text{tree}) *}_{ii \underline{0}}=\tilde{c}^{(\text{tree})}_{ii \underline{0}}$.
That means the coefficient tensors are real and do not involve any $\varepsilon$-tensors.

Using the above projection we obtain the following relevant coefficient tensors after expressing the scalar products of four momenta via independent Mandelstam variables ($u+s+t=4m^2$):
\begin{align}
 c^{(\text{tree-}s)}_{\underline{000}}s&=4m^4-\left(s+2t\right)m^2\,,&   
 c^{(\text{tree-}s)}_{\underline{110}}s&=m^4-t m^2+\tfrac{1}{8}s^2+\tfrac{1}{4}st+\tfrac{1}{4}t^2\,,\nonumber\\
 c^{(\text{tree-}s)}_{\underline{330}}s&=-\tfrac{1}{2}sm^2+\tfrac{1}{8}s^2+\tfrac{1}{4}st\,,\nonumber\\
 c^{(\text{tree-}t)}_{\underline{000}}t&=-2m^4+\left(\tfrac{3}{2}s-t\right)m^2-\tfrac{1}{4}s^2\,,&   
 c^{(\text{tree-}t)}_{\underline{110}}t&=-\tfrac{1}{2}m^4+\tfrac{1}{4}s t+\tfrac{1}{8}s^2+\tfrac{1}{8}t^2\,,\nonumber\\
 c^{(\text{tree-}t)}_{\underline{330}}t&=\tfrac{3}{2}m^4-\left(s+\tfrac{1}{2}t\right)m^2 +\tfrac{1}{4}st+\tfrac{1}{8}s^2+\tfrac{1}{8}t^2\,,&
 c^{(\text{tree-}t)}_{\underline{440}}t&=-\tfrac{1}{2}sm^2+\tfrac{1}{4}s^2\,. 
\end{align}
Using the above squaring formula we recover the well known unpolarized amplitude square result (without the usual initial state spin average normalization of $\tfrac{1}{4}$)
\begin{align}
 |\mathcal{M}^{\text{tree}}|^2 = 4^2\bigg[&4\bigg(\frac{1}{s^2}+\frac{1}{t^2}-\frac{1}{s t}\bigg)m^4-4\bigg(\frac{t}{s^2}+\frac{s}{t^2}\bigg)m^2\nonumber\\
 &+\frac{s^2}{t^2}+\frac{t^2}{s^2}+2\bigg(\frac{s}{t}+\frac{t}{s}\bigg)+3 \bigg]\,.
\end{align}

Another way to obtain the spin averaged matrix element square above is given by
\begin{align}
 |\mathcal{M}^{\text{tree}}|^2 = \frac{1}{(2 m)^4}\mathcal{M}^{\text{tree} *} \cdot \mathcal{M}^{\text{tree}}= d_s^2\sum\limits_{i j a} \mathcal{C}^{(\text{tree})*}_{ij a} \cdot \mathcal{C}^{(\text{tree})}_{ji a}\,.
\end{align}
To show that the above equation is true one just needs to recall that the projectors on particle and anti-particle states are given by
\begin{align}
 P^{\pm}_i= \frac{\slashed p_i \pm m}{2m} =(P^{\pm}_i)^2\,.
\end{align}
With the given formula it is straight forward to see that one can use the projector property for each external leg in order to reduce the amplitude square of $\mathcal{M}^{\text{tree}}$
to the product of $\mathcal{M}^{\text{tree}}$ with $\widetilde{\mathcal{M}}^{\text{tree}}$ times the prefactor $(2 m)^4$.
The technical disadvantage of the second way to calculate the matrix element square is that the summation over the (tensor) coefficient contractions is not restricted anymore to the simple $a=\underline{0}$ singlet subspace,
because amplitude and co-amplitude contain four momenta which can serve as source (and sink) of vector indices.
In fact we need to take into account all tensor coefficients and thus we cannot use the simplified coefficient basis $c^{(\text{tree})}$ (which in principle was only defined on the singlet subspace),
but use the full tensor basis\footnote{At this point we refrain from printing the lengthy result for all coefficients $\mathcal{C}^{(\text{tree})}$ here and just provide it in an electronic format.} $\mathcal{C}^{(\text{tree})}$.
However, performing the calculation along the second way gives a strong independent check of the used decomposition.
For example it is clear that all negative powers of $m$ have to vanish in the final summation which leads to a non-trivial cancellation 
of terms originating from different sub-blocks.
Besides the consistency checks provided in the electronic example programs,
we used the basis in order to check an independent result for the one-loop times one-loop box amplitude square contribution appearing at the two loop level in the Bhabha high energy scattering cross section at leading logarithmic order (subleading in the small electron mass expansion)~\cite{Penin:2016wiw}.

We would like to state some comments of technical nature concerning the presented method.
From the explicit definition of the projectors it becomes clear, 
that it can hardly be applied during any hand calculation.
The complexity of the appearing tensor expressions is too complicated and thus
they have to be treated within a computer algebra system\footnote{Which poses no runtime problem for today's computers.}.
For example compared to the method of helicity amplitudes it does not provide any simplification
of expressions in intermediate steps on amplitude level.
This is of course the case because we explicitly keep the full information of the amplitude
and do not require any specific on-shell conditions, 
which can be understood as a projection on a very specific subset of information contained in the (in general) unphysical amplitude.
However, the benefit of the method is that it works independent of the underlying kinematics and thus allows to obtain (off-shell) results
with increased re-use value. 
Further it gurantees a simple amplitude square calculation for arbitrary and thus even yet unknown amplitudes
relying only on the well developed Dirac trace technology.

\section{Summary and Conclusion}
We have presented $s$- and $t$-channel tensor bases which enable to project spin algebra dependent four fermion amplitudes in four dimenions
onto purely bosonic coefficient tensors.
The bases are formulated independently of any kinematics and solely rely on the four-dimensional Lorentz symmetry,
which allows to decompose the involved tensors on the subspaces of irreducible representations.
The general Fierz transformation relating the $s$- and $t$-channel basis tensors with each other have been obtained
in block diagonal form and can be used to related any obtained tensors coefficient in $s$-channel basis with the one in $t$-channel and vice versa.
Further any amplitude square can be evaluated by a contraction of the contributing tensor coefficients within the different irreducible subspaces
in block diagonal form.
The introduced bases retain the full (off-shell) amplitude information while removing any spin algebra objects on amplitude level
and thus allows to avoid more involved calculations on amplitude square level.
For the reader's convenience we provide explicit example implementations in electonic form.

\section*{Acknowledgements}
We would like to thank Alexander Penin for carefully reading the manuscript.

\newpage

\appendix
\section{Invariance Condition and Schouten Identies}\label{AppendixICSI}
\begin{figure}
\centering
\begin{tabular}{cc}
 \vcenteredhbox{$T\,=$} \vcenteredhbox{\includegraphics[scale=0.6]{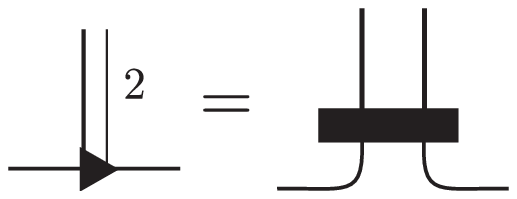}}&\vcenteredhbox{\includegraphics[scale=0.4]{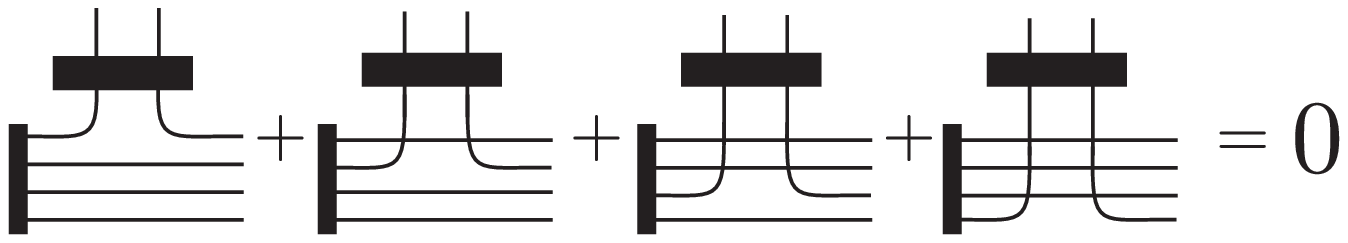}}\\
 (a) & (b)\\
 \multicolumn{2}{c}{\vcenteredhbox{\includegraphics[scale=0.4]{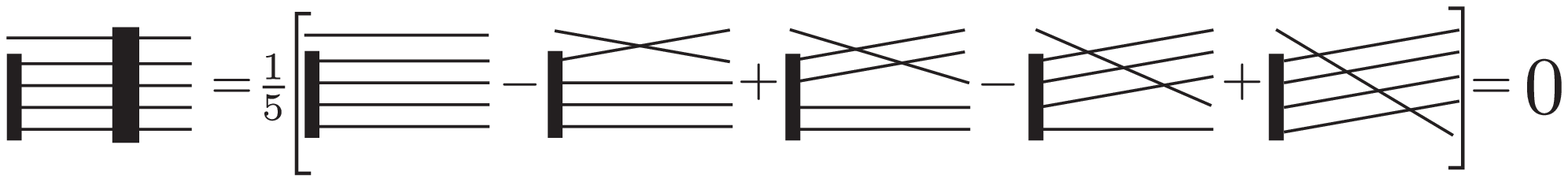}}}\\
 \multicolumn{2}{c}{(c)}\\
 \multicolumn{2}{c}{\vcenteredhbox{\includegraphics[scale=0.4]{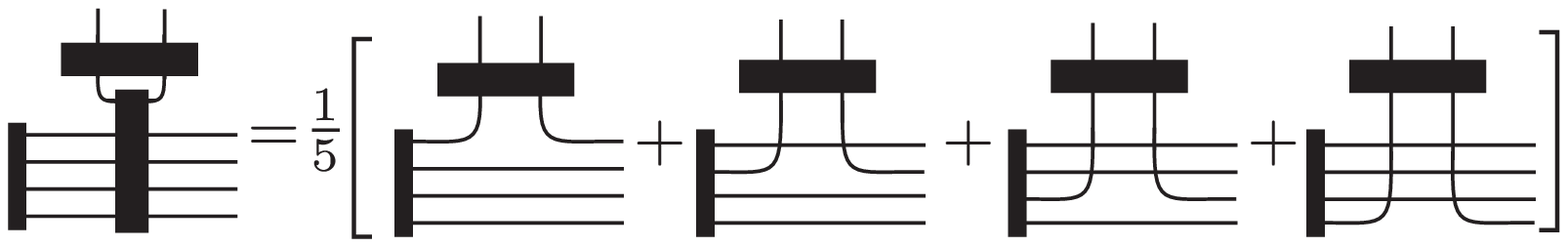}}}\\
 \multicolumn{2}{c}{(d)}
\end{tabular}
\caption{(a) Bird track notation of a generator in the vector representation. (b) Invariance condition for $\epsilon$-tensor. (c) Schouten identity.\label{FIG:lorentz_generator} (d) Relation between Schouten identity and invariance condition for $\epsilon$-tensor.}
\end{figure}

The non-trivally vanishing expressions in our case involve a direct product of one $\delta$- and one $\epsilon$-tensor for which not all combinations are linear independent.
For example the $\epsilon$-tensor fulfills the invariance condition:
\begin{align}
 \epsilon^{\mu \nu \rho \sigma} G(w)_{\mu}^{\mu^{\prime}}G(w)_{\nu}^{\nu^{\prime}}G(w)_{\rho}^{\rho^{\prime}}G(w)_{\sigma}^{\sigma^{\prime}}=\epsilon^{\mu^{\prime} \nu^{\prime} \rho^{\prime} \sigma^{\prime}}\,,
\end{align}
where we have for a group element:  
\begin{align}
 G(w)_{\mu}^{\mu^{\prime}}= \left[e^{i T^aw_a}\right]_{\mu}^{\mu^{\prime}}\,.
\end{align}
The $T^a$ are the generators of the group in the vector representation (see Fig.~\ref{FIG:lorentz_generator} (a) for a bird track notation) and $a$ is here an adjoint index.
Because the invariance condition holds for any group element, it must hold for the case where the arbitrary parameter $w^a$ are small and we can expand the exponent in Taylor-series.
Here we find in the linear order the non-trivial invariance condition:
\begin{align}
 (T^a)^{\mu^{\prime}}_{\mu}\epsilon^{\mu \nu^{\prime} \rho^{\prime} \sigma^{\prime}} +(T^a)^{\nu^{\prime}}_{\nu}\epsilon^{\mu^{\prime} \nu \rho^{\prime} \sigma^{\prime}} +(T^a)^{\rho^{\prime}}_{\rho}\epsilon^{\mu^{\prime} \nu^{\prime} \rho \sigma^{\prime}}+(T^a)^{\sigma^{\prime}}_{\sigma}\epsilon^{\mu^{\prime} \nu^{\prime} \rho^{\prime} \sigma}=0\,.
\end{align}
The explicit bird track notation in Fig.~\ref{FIG:lorentz_generator} (b) shows that the invariance condition requires a non-trivial relation between terms which are direct products of one $\epsilon$ and one $\delta$-tensor.
The non-trival behavior follows from the fact that one cannot have more than four fully antisymmetric vector indices in four dimensions.
Writing down a tensor with more than four fully antisymmetric indices in four dimensions yields zero.
This is also known as Schouten identity which is displayed in  Fig.~\ref{FIG:lorentz_generator} (c).
In Fig.~\ref{FIG:lorentz_generator} (d) we show that the anti-symmetric superposition of two Schouten identities yields the invariance condition of the $\epsilon$ tensor.

One can show that non-trivially vanishing tensor are indeed zero for e.g. working in an explicit component representation,
checking all components of the tensor.
However, this only allows one to check if a tensor is indeed zero,
but it does not automatically reduce a given analytic expression into a minimal set of tensors.
In turn this means that any explicit representation of an expressions is not uniquely determined and most of the time will contain more terms than needed.

In order to reduce all tensors of the type $E^{\mu_1 \mu_2 \mu_3 \mu_4 \mu_5 \mu_6}=g^{\mu_1 \mu_2}\tilde{\varepsilon}^{\mu_3 \mu_4 \mu_5 \mu_6}$ to a minimal set,
one first has to determine the number of linear independent tensors.
The Schouten identity in Fig.~\ref{FIG:lorentz_generator} (c) shows that one can eliminate one $E$-tensor in favour of four others.
All of the five tensors have the same Lorentz index $\mu_1$ (the upper left one in the diagram) in the argument of the metric tensor $g$.
Thus in general one can write down six different Schouten identities.
For a unique definition of the $E$-tensor we can use the first and second index $\mu_1$ and $\mu_2$ appearing in the metric tensor and thus we have $6\cdot5/2=15$ different tensor.
We can label them $E_{ij}=E_{ji}$ where $i$ and $j$ stand for the first $\mu_i$ and second index $\mu_j$ appearing in the metric tensor $g$.
That means each tensor appears in two different Schouten identities and we can only eliminate $5$ of the $15$ tensors using $5$ Schouten identities.
A convenient choice is to eliminate $E_{i i+1}$ for $i<6$.
Assuming that the remaining indices in the $\tilde{\varepsilon}$ tensor are to be ordered with increasing $k$ of the index $\mu_k$, meaning $E_{1 2}=E^{\mu_1 \mu_2 \mu_3 \mu_4 \mu_5 \mu_6}$
we obtain the elimination identities
\begin{align}
E_{5  6} =& -E_{1  6} + E_{2  6} - E_{3  6} + E_{4  6}\,,\nonumber\\ 
E_{4  5} =& +E_{1  5} - E_{1  6} - E_{2  5} + E_{2  6} + E_{3  5} - E_{3  6} + E_{4  6}\,,\nonumber\\ 
E_{3  4} =& -E_{1  4} + E_{1  5} - E_{1  6} + E_{2  4} - E_{2  5} + E_{2  6} + E_{3  5} - E_{3  6}\,,\nonumber\\ 
E_{2  3} =& +E_{1  3} - E_{1  4} + E_{1  5} - E_{1  6} + E_{2  4} - E_{2  5} + E_{2  6}\,,\nonumber\\ 
E_{1  2} =& +E_{1  3} - E_{1  4} + E_{1  5} - E_{1  6}\,.
\end{align}

\section{Electronic Implementation and Results}\label{AppendixElectronics}
For the reader's convenience we provide electronic result files allowing to apply the introduced projectors onto the basis within \textsc{FORM} 
and carrying out the Fierz transformations in \textsc{FORM} or \textsc{Mathematica}.
We also provide two example \textsc{FORM} programs demonstrating how to use the implemented routines and results.
All provided files are listed in Tab.~\ref{TAB:eFileListing}.
\begin{table}
\centering
\begin{tabular}{|l|l|l|} \hline
File name& \multicolumn{2}{l|}{Description} \\ \hline\hline
\verb$fierzing.frm$& \multicolumn{2}{p{6cm}|}{\raggedright \footnotesize Main \textsc{FORM} fold package providing projection and Fierz transformation routines.} \\ \hline
\verb$fierzing.inc$& \multicolumn{2}{p{6cm}|}{\raggedright \footnotesize Contains the Fierz transformations in \textsc{FORM} syntax.} \\ \hline
\verb$fierzing.m$& \multicolumn{2}{p{6cm}|}{\raggedright \footnotesize Contains the Fierz transformations in \textsc{Mathematica} syntax.} \\ \hline
\verb$bhabhatree.frm$& \multicolumn{2}{p{6cm}|}{\raggedright \footnotesize Example \textsc{FORM} program containing the Bhabha scattering amplitude square at tree level.} \\ \hline
\verb$examples.frm$& \multicolumn{2}{p{6cm}|}{\raggedright \footnotesize Example \textsc{FORM} program containing generic amplitude square calculation involving $\gamma_5$.} \\ \hline
\end{tabular}
\caption{Listing of provided electronic files included in this publication. \label{TAB:eFileListing}
}
\end{table}
In order to be able to relate the notation in the electronic files to the one in this publication we provide a small dictionary in  Tab.~\ref{TAB:eDictionary}. 
\begin{table}
\centering
\begin{tabular}{|c c|c c|} \hline
\verb$d_(mu1,mu2)$ & $g^{\mu_1 \mu_2}$ & \verb$e_(mu1,mu2,mu3,mu4)$ & $\sqrt{4!}\tilde\varepsilon^{\mu_1 \mu_2 \mu_3 \mu_4}$\\ \hline
\verb$ss`i'`j'to`a'$  & $|i,j;a\rangle_{(s)}$ & \verb$tt`i'`j'to`a'$ & $|i,j;a\rangle_{(t)}$\\ \hline
\verb$sqr2$ & $\sqrt{2}$  & & \\ \hline
\end{tabular}
\caption{Definition of the used electronic notation. \label{TAB:eDictionary}}
\end{table}
We use the notation $n^+=\,$\verb$nS$ and $n^-=\,$\verb$nA$ in order to denote the irreps with parity.
The definition of the \verb$` '$ operation can be understood just looking at some explicit examples.
We have $|\underline{2},\underline{2};\underline{4}\rangle_{(s)}=|\underline{2},\underline{2};1_a\rangle_{(s)}=$~\verb$ss66to1a$, $|\underline{4},\underline{1};9\rangle_{(t)}=$~\verb$tt1A4Sto9$, $|\underline{4},\underline{0};\underline{4}\rangle_{(t)}=|\underline{4},\underline{0};1\rangle_{(t)}=$~\verb$tt1A1Sto1$.
In the electronic files we do not use the bilinear notation for the index $a$. But we just state the representation label $\mathcal{R}_a$ appearing in the decomposition.
This fixes the basis tensor uniquely, although it does not display any parity information.

In order to keep the output more readable we keep the Lorentz vector indices implicit inside the symbol.

We have the following convention.
In the $s$-channel setup the incoming particles are connected to chain one  (\verb$C1$) carrying bilinear index $i$ with possible Lorentz indices $\sigma_1$ and $\sigma_2$.
Only in the case $i=\underline{2}$ both Lorentz indices are present.
In case of $i=\underline{1}$ and $i=\underline{3}$ only $\sigma_1$ is active. 
In the remaining (pseudo) scalar case no Lorentz index is present.
The outgoing particles are connected to chain two (\verb$C2$) carrying bilinear index $j$ with possible Lorentz indices $\sigma_3$ and $\sigma_4$.
Only in the case $j=\underline{2}$ both Lorentz indices are present.
In case of $j=\underline{1}$ and $j=\underline{3}$ only $\sigma_3$ is active. 
In the remaining (pseudo) scalar case no Lorentz index is present.

In the $t$-channel setup the incoming and outgoing (anti-)fermion couples to the same Dirac chain \verb$(C4)$\verb$C3$ with index $(j) i$ carrying the optional Lorentz indices ($\nu_3$ and $\nu_4$) $\nu_1$ and $\nu_2$.
The presence of the stated Lorentz indices is determined in the same way like for an $s$-channel basis element.

For the bra basis elements  $\langle i,j;a|$ we do not introduce a separate notation. 
It arises from the program context, if one is dealing with bras or kets.

In the file \verb$fierzing.inc$ we store all Fierz relations required to transform any $s$-channel basis tensors into $t$-channel ones and vice versa.
For each basis tensors it contains an explicit expression with name \verb$ss`i'`j'to`a'Rule$/\verb$tt`i'`j'to`a'Rule$ 
which contains the respective basis tensor $|i,j;a\rangle_{(s)}$/$|i,j;a\rangle_{(t)}$ expressed in terms of the $t$/$s$-channel basis tensors.
That means the replacement \verb$ss`i'`j'to`a'$ $\rightarrow$ \verb$ss`i'`j'to`a'Rule$ will do the Fierz transformation of the $s$-channel basis element $|i,j;a\rangle_{(s)}$
into the respective $t$-channel basis elements $|k,l;b\rangle_{(t)}$.

For example we can find the definition
\begin{lstlisting}[frame=single]
 L  ss4S1Ato4Rule = (
 + sqr2^-1*tt64Ato4b*(- 1/4*d_(nu1,nu3)*d_(nu2,sigma1)+1/4*d_(nu1,sigma1)*d_(nu2,nu3))
 + sqr2^-1*tt64Sto4a*( 1/4*e_(nu1,nu2,nu3,sigma1) )
 + sqr2^-1*tt4A6to4b*( 1/4*d_(nu1,nu3)*d_(nu4,sigma1)-1/4*d_(nu1,nu4)*d_(nu3,sigma1))
 + sqr2^-1*tt4S6to4a*( 1/4*e_(nu1,nu3,nu4,sigma1) )
 + tt4A1Sto4*(  - 1/4*d_(nu1,sigma1) )
 + tt4S1Ato4*( 1/4*d_(nu1,sigma1) )
 + tt1S4Ato4*( 1/4*d_(nu3,sigma1) )
 + tt1A4Sto4*( 1/4*d_(nu3,sigma1) )
);
\end{lstlisting}
which corresponds to ($\hat{\varepsilon}=\sqrt{4!}\tilde{\varepsilon}$)
\begin{align}
 |\underline{1},\underline{4};4\rangle_{(s)}^{\sigma_1}=&+\tfrac{1}{\sqrt{2}}|\underline{2},\underline{3};4_b\rangle_{(t)}^{\nu_1 \nu_2 \nu_3}\Big(-\tfrac{1}{4}g_{\nu_1\nu_3}g_{\nu_2}^{\sigma_1}+\tfrac{1}{4}g_{\nu_2\nu_3}g_{\nu_1}^{\sigma_1}\Big)\nonumber\\
 &+\tfrac{1}{\sqrt{2}}|\underline{2},\underline{1};4_a\rangle_{(t)}^{\nu_1 \nu_2 \nu_3}\Big(\tfrac{1}{4} \hat{\varepsilon}_{\nu_1\nu_2 \nu_3}^{\phantom{\nu_1\nu_2 \nu_3}\sigma_1}\Big)\nonumber\\
 &+\tfrac{1}{\sqrt{2}}|\underline{3},\underline{2};4_b\rangle_{(t)}^{\nu_1 \nu_3 \nu_4}\Big(\tfrac{1}{4}g_{\nu_1\nu_3}g_{\nu_4}^{\sigma_1}-\tfrac{1}{4}g_{\nu_1\nu_4}g_{\nu_3}^{\sigma_1}\Big)\nonumber\\
 &+\tfrac{1}{\sqrt{2}}|\underline{1},\underline{2};4_a\rangle_{(t)}^{\nu_1 \nu_3 \nu_4}\Big(\tfrac{1}{4} \hat{\varepsilon}_{\nu_1\nu_3 \nu_4}^{\phantom{\nu_1\nu_3 \nu_4}\sigma_1}\Big)\nonumber\\
 &+|\underline{3},\underline{0};4\rangle_{(t)}^{\nu_1}\Big(-\tfrac{1}{4} g_{\nu_1}^{\sigma_1}\Big)\nonumber\\
 &+|\underline{1},\underline{4};4\rangle_{(t)}^{\nu_1}\Big(\tfrac{1}{4} g_{\nu_1}^{\sigma_1}\Big)\nonumber\\
 &+|\underline{0},\underline{3};4\rangle_{(t)}^{\nu_3}\Big(\tfrac{1}{4} g_{\nu_3}^{\sigma_1}\Big)\nonumber\\
 &+|\underline{4},\underline{1};4\rangle_{(t)}^{\nu_3}\Big(\tfrac{1}{4} g_{\nu_3}^{\sigma_1}\Big)\,.
\end{align}



\printindex

\end{document}